\RequirePackage{fix-cm}
\documentclass[reqno]{amsproc}
\usepackage{amssymb}

\usepackage[citation-order,nobysame]{amsrefs}
\usepackage{xyzbib}

\usepackage{mathtools}

\usepackage{hyperref}

\usepackage{tikz}
\usetikzlibrary{decorations.pathreplacing}

\mathtoolsset{showonlyrefs}
\numberwithin{equation}{section}
\let\cite=\cites

\DeclareMathOperator*{\Asym}{\textrm{Asym}}

\let\leq=\leqslant

\newcommand{\rmi}{\mathrm{i}}
\newcommand{\rmd}{\mathrm{d}}
\newcommand{\eps}{\varepsilon}

\newcommand{\ot}{\mathop{\otimes}}
\newcommand{\caM}{\mathcal{M}}
\newcommand{\caN}{\mathcal{N}}

\newcommand{\bra}[1]{\langle #1 \rvert}
\newcommand{\ket}[1]{\lvert #1 \rangle}

\newcommand{\psitop}[1]{Z^{\mathrm{top}}_{#1}}
\newcommand{\psibot}[1]{Z^{\mathrm{bot}}_{#1}}
\newcommand{\subtop}{_{\mathrm{top}}}
\newcommand{\subbot}{_{\mathrm{bot}}}
\newcommand{\rbar}{\bar{r}}
\newcommand{\Clambdas}{{C_{\{\lambda\}}}}
\newcommand{\Cws}{{C_{\{1/w\}}}}
\usepackage{currfile}
\begin{document}


\title[integral representations for nonlocal correlation
  functions]{Six-vertex model on a finite lattice: integral
  representations for nonlocal correlation functions}

\author{F. Colomo}
\address{INFN, Sezione di Firenze\\
Via G. Sansone 1, I-50019 Sesto Fiorentino (FI), Italy}
\email{colomo@fi.infn.it}

\author{G. Di Giulio}
\address{SISSA and INFN Sezione di Trieste\\
Via Bonomea 265, I-34136 Trieste, Italy}
\email{gdigiuli@sissa.it}
  
\author{A. G. Pronko}
\address{Steklov Mathematical Institute, 
  Fontanka 27, 191023 Saint Petersburg, Russia  and
  Theoretical Physics Department, Saint Petersburg State University,
Ulyanovskaya str. 1, Peterhof, Saint Petersburg, 198504, Russia}
\email{agp@pdmi.ras.ru}

\begin{abstract}  

We consider the problem of calculation of correlation functions in the
six-vertex model with domain wall boundary conditions.  To this aim,
we formulate the model as a scalar product of off-shell Bethe states,
and, by applying the quantum inverse scattering method, we derive
three different integral representations for these states.  By
suitably combining such representations, and using certain
antisymmetrization relation in two sets of variables, it is possible
to derive integral representations for various correlation
functions. In particular, focusing on the emptiness formation
probability, besides reproducing the known result, obtained by other
means elsewhere, we provide a new one. By construction, the two
representations differ in the number of integrations and their
equivalence is related to a hierarchy of highly nontrivial identities.

\end{abstract}

\maketitle
\setcounter{tocdepth}{2}
\tableofcontents
\section{Introduction}

The theory of quantum integrable models can be seen as developing
mainly in two directions (see, e.g., \cite{KBI-93}).  One is related
to various (algebraic, analytic and geometric) aspects of
integrability (such as Bethe ansatz, Yang-Baxter relation, Baxter T-Q
equation, Quantum Groups, etc) of quantum models and mostly deals with
the problem of diagonalization of quantum integrals of motion
(transfer matrices).  The other is related to applications (in gauge
and string theories, condensed matters, algebraic combinatorics,
probability, etc.)  and deals mostly with the problem of evaluation of
correlation functions.  The latter uses heavily results from the
former, but despite significant progress of the theory in general, it
still provides numerous challenging problems.  One of them concerns
the calculation of correlation functions of models with broken
translational invariance (e.g., due to the boundary conditions),
possibly in cases where keeping finite the size of the system is of
importance for extracting interesting (physical or mathematical)
information.

A prototypical example is provided by the six-vertex model with domain
wall boundary conditions. Current interest in the model is mostly
motivated by the occurrence of phase separation
\cite{KZj-00,Zj-00,SZ-04,PR-07}, which recently triggered a number of
numerical studies \cite{LKRV-18,KL-17,BR-20} and analytical results
\cite{CS-16,RS-17,KRS-20,KP-20}. The model is also of relevance for
quantum quenches in the closely related Heisenberg XXZ quantum spin
chain \cite{ADSV-16,S-17,CDV-18,S-20}, and for $\mathcal{N}=4$
super-Yang-Mills theory \cite{F-11,K-12,JKKS-16}.  As for
correlation functions of the six-vertex model in the bulk (or with
periodic boundary conditions), many notable important results were
obtained for the XXZ chain (see, among others, papers
\cite{KMST-02,BKNS-02,BKS-03,GKS-04,BJMST-06,KKMST-09,JMS-09,MS-19,K-19}
and references therein). Although these results cannot be directly
applied to the case of domain wall boundary conditions, some aspects,
such as multiple integral representations, indeed do prove useful.

Historically, correlation functions of the six-vertex model with
domain wall boundary conditions were first studied close to the
boundary, where the problem notably simplifies
\cite{BKZ-02,BPZ-02,FP-04,CP-05c,M-11}. Some results and techniques
developed in these studies allowed to evaluate a bulk correlation
function, the so-called emptiness formation probability, as a multiple
integral \cite{CP-07b,CPS-16}.  Subsequent study of such
integral representation made it possible to derive an exact analytic
expression for the spatial curve separating ferroelectric order from
disorder (the so-called `arctic' curve)
\cite{CP-07a,CP-08,CP-09,CPZj-10}.

In \cite{CP-12}, in order to extend previous progress with the
emptiness formation probability to other correlation functions, a
method for their systematic calculation was proposed. A
specific nonlocal correlation function, named row configuration
probability, was introduced, in terms of which various other (nonlocal
and local) correlation functions can in principle be obtained by
suitable summations over position parameters. The row configuration
probability can be viewed as a product of two factors, which are in
fact components of off-shell Bethe states. These states are
complementary to each other in the sense that they are build over
different pseudo-vacuum states (all-spins-up and all-spins-down);
their scalar product is exactly the partition function. The main
observation in \cite{CP-12} is that these two factors can be
represented as multiple integrals involving the \emph{same number} of
integrations.

The main difficulty of the proposed method concerns the possibility of
performing suitable summations and integrations to simplify the
resulting expressions. As already noted in \cite{CP-12}, to reproduce
the previously obtained integral representation for the emptiness
formation probability, one has to deal with a nontrivial problem of
antisymmetrization over two sets of integration variables. In this
respect, the existence of some useful antisymmetrization identity
allows to show that indeed the proposed approach can be useful in
practice \cite{CCP-19}.

In the present paper, we overview the method of \cite{CP-12,CCP-19}
and provide some improvements which allow  us to obtain further
results.  We start by formulating the model as a scalar product of
off-shell Bethe states; applying the Quantum Inverse Scattering Method
(QISM) \cite{TF-79} (see also \cite{KBI-93} and references therein),
we derive three constructively different integral representations for
the components of the off-shell Bethe states.

Next, we show how such representations can be combined to build
various local or nonlocal correlation functions.  In particular,
focusing on the emptiness formation probability, we combine two out of
the three available representations and show how the problem of
evaluating some intricate multiple sums and integrals can be tackled
so that one can recover the multiple integral representation first
derived in \cite{CP-07b}.  The antisymmetrization relation proven in
\cite{CCP-19} plays a crucial role in this alternative derivation.

Finally, by combining a different pair of representations for the
components of the off-shell Bethe states, we derive an alternative
multiple integral representation for the emptiness formation
probability.  The existence of two essentially different integral
representations for the same object leads to a hierarchy of identities
involving the (generating function of the) one-point boundary
correlation function.  Further study of such identities is required
for a full understanding of their implications.

The paper is organized as follows. In the next section, after giving
some definitions and notations, we sketch the strategy of our
derivation, and set up QISM in the context of the considered problem.
The core calculation of the components of the off-shell Bethe states
for the inhomogeneous model is contained in section 3.  In section 4
we perform the homogeneous limit, and obtain representations in terms
of multiple integrals.  In section 5, starting from a suitable
combination of two such representations, we reproduce the previously
obtained integral representation for the emptiness probability. In
section 6 we show that the same procedure, when applied to a different
combination of components of off-shell Bethe states, leads to an
alternative and essentially different integral representation for the
emptiness formation probability.

\section{The model}

In this section we define the model, introduce some nonlocal
correlation functions of interest, and formulate the problem in the
framework the QISM.

\subsection{The partition function}

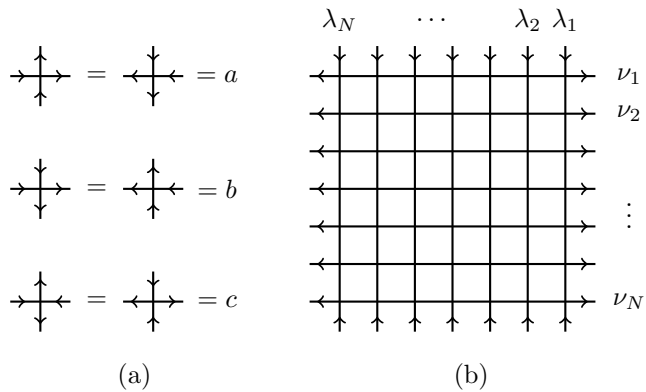
\begin{figure}
\centering

\begin{tikzpicture}[scale=.5]
\draw [thick] (0.2,7)--(1.8,7);
\draw [thick] (1,6.2)--(1,7.8);
\draw [thick] [->] (0.5,7)--(0.6,7);
\draw [thick] [->] (1.5,7)--(1.6,7);
\draw [thick] [->] (1,7.5)--(1,7.6);
\draw [thick] [->] (1,6.5)--(1,6.6);
\node at (2.5,7) {$=$};
\draw [thick] (3.2,7)--(4.8,7);
\draw [thick] (4,6.2)--(4,7.8);
\draw [thick] [->] (3.5,7)--(3.4,7);
\draw [thick] [->] (4.5,7)--(4.4,7);
\draw [thick] [->] (4,7.5)--(4,7.4);
\draw [thick] [->] (4,6.5)--(4,6.4);
\node at (5.7,7) {$=a$};
\draw [thick] (0.2,4)--(1.8,4);
\draw [thick] (1,3.2)--(1,4.8);
\draw [thick] [->] (0.5,4)--(0.6,4);
\draw [thick] [->] (1.5,4)--(1.6,4);
\draw [thick] [->] (1,4.5)--(1,4.4);
\draw [thick] [->] (1,3.5)--(1,3.4);
\node at (2.5,4) {$=$};
\draw [thick] (3.2,4)--(4.8,4);
\draw [thick] (4,3.2)--(4,4.8);
\draw [thick] [->] (3.5,4)--(3.4,4);
\draw [thick] [->] (4.5,4)--(4.4,4);
\draw [thick] [->] (4,4.5)--(4,4.6);
\draw [thick] [->] (4,3.5)--(4,3.6);
\node at (5.7,4) {$=b$};
\draw [thick] (0.2,1)--(1.8,1);
\draw [thick] (1,0.2)--(1,1.8);
\draw [thick] [->] (0.5,1)--(0.6,1);
\draw [thick] [->] (1.5,1)--(1.4,1);
\draw [thick] [->] (1,1.5)--(1,1.6);
\draw [thick] [->] (1,0.5)--(1,0.4);
\node at (2.5,1) {$=$};
\draw [thick] (3.2,1)--(4.8,1);
\draw [thick] (4,0.2)--(4,1.8);
\draw [thick] [->] (3.5,1)--(3.4,1);
\draw [thick] [->] (4.5,1)--(4.6,1);
\draw [thick] [->] (4,1.5)--(4,1.4);
\draw [thick] [->] (4,0.5)--(4,0.6);
\node at (5.7,1) {$=c$};
\node at (3.5,-1) {(a)};
\end{tikzpicture}
\qquad
\begin{tikzpicture}[scale=.5]
\draw [thick] (0.2,1)--(7.8,1);
\draw [thick] (0.2,2)--(7.8,2);
\draw [thick] (0.2,3)--(7.8,3);
\draw [thick] (0.2,4)--(7.8,4);
\draw [thick] (0.2,5)--(7.8,5);
\draw [thick] (0.2,6)--(7.8,6);
\draw [thick] (0.2,7)--(7.8,7);
\draw [thick] (1,0.2)--(1,7.8);
\draw [thick] (2,0.2)--(2,7.8);
\draw [thick] (3,0.2)--(3,7.8);
\draw [thick] (4,0.2)--(4,7.8);
\draw [thick] (5,0.2)--(5,7.8);
\draw [thick] (6,0.2)--(6,7.8);
\draw [thick] (7,0.2)--(7,7.8);
\draw [thick] [->] (.5,1)--(.4,1);
\draw [thick] [->] (.5,2)--(.4,2);
\draw [thick] [->] (.5,3)--(.4,3);
\draw [thick] [->] (.5,4)--(.4,4);
\draw [thick] [->] (.5,5)--(.4,5);
\draw [thick] [->] (.5,6)--(.4,6);
\draw [thick] [->] (.5,7)--(.4,7);
\draw [thick] [->] (1,7.5)--(1,7.4);
\draw [thick] [->] (2,7.5)--(2,7.4);
\draw [thick] [->] (3,7.5)--(3,7.4);
\draw [thick] [->] (4,7.5)--(4,7.4);
\draw [thick] [->] (5,7.5)--(5,7.4);
\draw [thick] [->] (6,7.5)--(6,7.4);
\draw [thick] [->] (7,7.5)--(7,7.4);
\draw [thick] [->] (7.5,1)--(7.6,1);
\draw [thick] [->] (7.5,2)--(7.6,2);
\draw [thick] [->] (7.5,3)--(7.6,3);
\draw [thick] [->] (7.5,4)--(7.6,4);
\draw [thick] [->] (7.5,5)--(7.6,5);
\draw [thick] [->] (7.5,6)--(7.6,6);
\draw [thick] [->] (7.5,7)--(7.6,7);
\draw [thick] [->] (1,.5)--(1,.6);
\draw [thick] [->] (2,.5)--(2,.6);
\draw [thick] [->] (3,.5)--(3,.6);
\draw [thick] [->] (4,.5)--(4,.6);
\draw [thick] [->] (5,.5)--(5,.6);
\draw [thick] [->] (6,.5)--(6,.6);
\draw [thick] [->] (7,.5)--(7,.6);
\node at (8.7,7) {$\nu_1$};
\node at (8.7,6) {$\nu_2$};
\node at (8.7,3.5) {$\vdots$};
\node at (8.7,1) {$\nu_N$};
\node at (7,8.5) {$\lambda_1$};
\node at (6,8.5) {$\lambda_2$};
\node at (3.5,8.5) {$\dots$};
\node at (1,8.5) {$\lambda_N$};
\node at (4.5,-1) {(b)};
\end{tikzpicture}

\caption{The model: (a) The six possible vertex configurations, and
  their weights; (b) The $N\times N$ square lattice with domain wall
  boundary conditions; also shown is the assigment of parameters
  $\lambda_1,\ldots,\lambda_N$ and $\nu_1,\ldots,\nu_N$ to the
  vertical and horizontal lines, respectively.}
\label{fig-6V-DW}
\end{figure}

The six-vertex model with domain wall boundary conditions is a model
of arrows lying on the edges of a square lattice with $N$ horizontal
and $N$ vertical lines.  Arrow configurations are constrained by the
`ice-rule', requiring each vertex to have the same number of incoming
and outgoing arrows \cite{B-82}.  Boltzmann weights are assigned to
the six possible vertex configurations of arrows allowed by the
ice-rule. With no loss of generality, we require the model to be
invariant under reversal of all arrows.  We thus have three distinct
Boltzmann weights, denoted $a$, $b$, $c$.  The domain wall boundary
conditions are defined as follows: all arrows on the left and right
boundaries are outgoing while all arrows on the top and bottom
boundaries are incoming, see figure \ref{fig-6V-DW}.

To use QISM for calculations we will consider the inhomogeneous
version of the model, in which the weights of the vertex being at the
intersection of $k$th horizontal line and $\alpha$th vertical line are
\cite{B-71}
\begin{equation}\label{abc-im}
a_{\alpha k}=a(\lambda_\alpha,\nu_k),\qquad
b_{\alpha k}=b(\lambda_\alpha,\nu_k),\qquad
c_{\alpha k}=c,
\end{equation}
where
\begin{equation}\label{abc}
a(\lambda,\nu)=\sin(\lambda-\nu+\eta),\qquad
b(\lambda,\nu)=\sin(\lambda-\nu-\eta),\qquad
c=\sin2\eta,
\end{equation}
and we enumerate vertical lines (labelled by Greek indices) from right
to left, and horizontal lines (labelled by Latin indices) from top to
bottom.  The parameters $\lambda_1,\dots,\lambda_N$ are assumed to be
all different; the same is assumed about $\nu_1,\dots,\nu_N$.  To obtain the
homogeneous model, after applying QISM, we set these parameters equal
within each set, $\lambda_\alpha=\lambda$ and $\nu_k=\nu$ where, with
no loss of generality, we can choose $\nu=0$.

The partition function of the inhomogeneous model is defined as the
sum over all possible configurations, each configuration being
assigned its Boltzmann weight, which is the product of all vertex
weights over the lattice,
\begin{equation}
Z_N= \sum_{\mathcal{C}}^{}
\prod_{\alpha=1}^{N}\prod_{k=1}^{N}w_{\alpha k}(C).
\end{equation}
Here $w_{\alpha k}(C)$ takes values $w_{\alpha k}(C)= a_{\alpha
  k},b_{\alpha k},c_{\alpha k}$, depending on the configuration
$C$. Because of \eqref{abc-im},
$Z_N=Z_N(\lambda_1,\dots,\lambda_N;\nu_1,\dots,\nu_N)$ where the
$\lambda$'s and $\nu$'s are regarded as `variables'; $\eta$ is
regarded as a parameter (having the meaning of a `coupling constant')
and it is often omitted in notations. In QISM the dependence on
$\lambda$'s and $\nu$'s play an important role (in particular, $Z_N$
is invariant under permutations within each set of variables).

The partition function may be written in determinantal form,  
as established by Izergin and Korepin in \cites{K-82,I-87,ICK-92}:
\begin{equation}\label{ZN}
Z_N=
\frac{\prod_{\alpha=1}^N \prod_{k=1}^N
a(\lambda_\alpha,\nu_k)b(\lambda_\alpha,\nu_k)}{
\prod_{1\leq\alpha<\beta\leq N}d(\lambda_\beta,\lambda_\alpha)
\prod_{1\leq j<k\leq N}d(\nu_j,\nu_k)}\,
\det\caM.
\end{equation}
Here $\caM$ is an $N$-by-$N$ matrix with entries
\begin{equation}\label{matT}
\caM_{\alpha k}=\varphi(\lambda_\alpha,\nu_k),\qquad
\varphi(\lambda,\nu)=\frac{c}{a(\lambda,\nu) b(\lambda,\nu)},
\end{equation}
while $a(\lambda,\nu)$, $b(\lambda,\nu)$ and $c$ are defined in
\eqref{abc}, and function $d(\lambda,\lambda')$, standing in the
prefactor of \eqref{ZN}, is
\begin{equation}\label{dfunc}
d(\lambda,\lambda'):=\sin(\lambda-\lambda').
\end{equation}
In the homogeneous limit, i.e., when $\lambda_\alpha=\lambda$ and
$\nu_k=0$, expression \eqref{ZN} becomes
\begin{equation}\label{ZNhom}
Z_N=\frac{[\sin(\lambda-\eta)\sin(\lambda+\eta)]^{N^2}}
{\prod_{n=1}^{N-1}(n!)^2}\, \det\caN,
\end{equation}
where the $N$-by-$N$ matrix $\caN$ has entries
\begin{equation}\label{varphi}
\caN_{\alpha k}=\partial_{\lambda}^{\alpha+k-2}
\varphi(\lambda),\qquad
\varphi(\lambda):=\varphi(\lambda,0)=
\frac{\sin 2\eta}{\sin(\lambda-\eta)\sin(\lambda+\eta)}.
\end{equation}
For a detailed proof of  \eqref{ZN} and \eqref{ZNhom}, see \cite{ICK-92}.
For an alternative derivation, see \cites{BPZ-02,CP-07b}.

\subsection{Off-shell Bethe states  and correlation functions}\label{sec.confprobs}

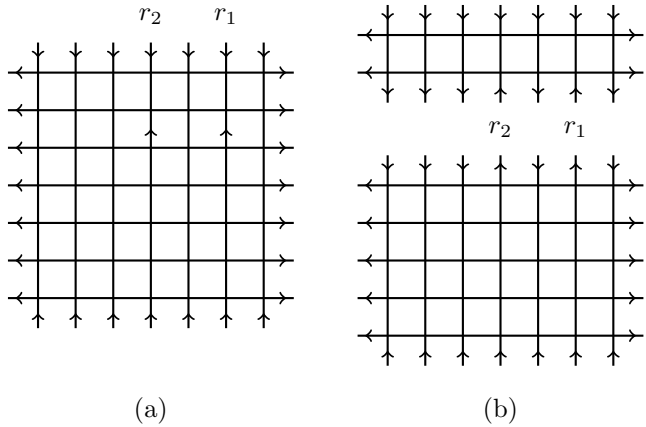
\begin{figure}
\centering

\begin{tikzpicture}[scale=.5]
\draw [thick] (0.2,1)--(7.8,1);
\draw [thick] (0.2,2)--(7.8,2);
\draw [thick] (0.2,3)--(7.8,3);
\draw [thick] (0.2,4)--(7.8,4);
\draw [thick] (0.2,5)--(7.8,5);
\draw [thick] (0.2,6)--(7.8,6);
\draw [thick] (0.2,7)--(7.8,7);
\draw [thick] (1,0.2)--(1,7.8);
\draw [thick] (2,0.2)--(2,7.8);
\draw [thick] (3,0.2)--(3,7.8);
\draw [thick] (4,0.2)--(4,7.8);
\draw [thick] (5,0.2)--(5,7.8);
\draw [thick] (6,0.2)--(6,7.8);
\draw [thick] (7,0.2)--(7,7.8);
\draw [thick] [->] (.5,1)--(.4,1);
\draw [thick] [->] (.5,2)--(.4,2);
\draw [thick] [->] (.5,3)--(.4,3);
\draw [thick] [->] (.5,4)--(.4,4);
\draw [thick] [->] (.5,5)--(.4,5);
\draw [thick] [->] (.5,6)--(.4,6);
\draw [thick] [->] (.5,7)--(.4,7);
\draw [thick] [->] (1,7.5)--(1,7.4);
\draw [thick] [->] (2,7.5)--(2,7.4);
\draw [thick] [->] (3,7.5)--(3,7.4);
\draw [thick] [->] (4,7.5)--(4,7.4);
\draw [thick] [->] (5,7.5)--(5,7.4);
\draw [thick] [->] (6,7.5)--(6,7.4);
\draw [thick] [->] (7,7.5)--(7,7.4);
\draw [thick] [->] (7.5,1)--(7.6,1);
\draw [thick] [->] (7.5,2)--(7.6,2);
\draw [thick] [->] (7.5,3)--(7.6,3);
\draw [thick] [->] (7.5,4)--(7.6,4);
\draw [thick] [->] (7.5,5)--(7.6,5);
\draw [thick] [->] (7.5,6)--(7.6,6);
\draw [thick] [->] (7.5,7)--(7.6,7);
\draw [thick] [->] (1,.5)--(1,.6);
\draw [thick] [->] (2,.5)--(2,.6);
\draw [thick] [->] (3,.5)--(3,.6);
\draw [thick] [->] (4,.5)--(4,.6);
\draw [thick] [->] (5,.5)--(5,.6);
\draw [thick] [->] (6,.5)--(6,.6);
\draw [thick] [->] (7,.5)--(7,.6);
\draw [thick] [->] (4,5.4)--(4,5.5);
\draw [thick] [->] (6,5.4)--(6,5.5);
\node at (4,8.5) {$r_2$};
\node at (6,8.5) {$r_1$};
\node at (4,-2) {(a)};
\end{tikzpicture}
\qquad
\begin{tikzpicture}[scale=.5]
\draw [thick] (0.2,0)--(7.8,0);
\draw [thick] (0.2,1)--(7.8,1);
\draw [thick] (0.2,2)--(7.8,2);
\draw [thick] (0.2,3)--(7.8,3);
\draw [thick] (0.2,4)--(7.8,4);
\draw [thick] (0.2,8)--(7.8,8);
\draw [thick] (0.2,7)--(7.8,7);
\draw [thick] (1,-.8)--(1,4.8);
\draw [thick] (2,-.8)--(2,4.8);
\draw [thick] (3,-.8)--(3,4.8);
\draw [thick] (4,-.8)--(4,4.8);
\draw [thick] (5,-.8)--(5,4.8);
\draw [thick] (6,-.8)--(6,4.8);
\draw [thick] (7,-.8)--(7,4.8);
\draw [thick] (1,6.2)--(1,8.8);
\draw [thick] (2,6.2)--(2,8.8);
\draw [thick] (3,6.2)--(3,8.8);
\draw [thick] (4,6.2)--(4,8.8);
\draw [thick] (5,6.2)--(5,8.8);
\draw [thick] (6,6.2)--(6,8.8);
\draw [thick] (7,6.2)--(7,8.8);
\draw [thick] [->] (.5,1)--(.4,1);
\draw [thick] [->] (.5,2)--(.4,2);
\draw [thick] [->] (.5,3)--(.4,3);
\draw [thick] [->] (.5,4)--(.4,4);
\draw [thick] [->] (.5,0)--(.4,0);
\draw [thick] [->] (.5,8)--(.4,8);
\draw [thick] [->] (.5,7)--(.4,7);
\draw [thick] [->] (1,8.5)--(1,8.4);
\draw [thick] [->] (2,8.5)--(2,8.4);
\draw [thick] [->] (3,8.5)--(3,8.4);
\draw [thick] [->] (4,8.5)--(4,8.4);
\draw [thick] [->] (5,8.5)--(5,8.4);
\draw [thick] [->] (6,8.5)--(6,8.4);
\draw [thick] [->] (7,8.5)--(7,8.4);
\draw [thick] [->] (7.5,1)--(7.6,1);
\draw [thick] [->] (7.5,2)--(7.6,2);
\draw [thick] [->] (7.5,3)--(7.6,3);
\draw [thick] [->] (7.5,4)--(7.6,4);
\draw [thick] [->] (7.5,0)--(7.6,0);
\draw [thick] [->] (7.5,8)--(7.6,8);
\draw [thick] [->] (7.5,7)--(7.6,7);
\draw [thick] [->] (1,-.5)--(1,-.4);
\draw [thick] [->] (2,-.5)--(2,-.4);
\draw [thick] [->] (3,-.5)--(3,-.4);
\draw [thick] [->] (4,-.5)--(4,-.4);
\draw [thick] [->] (5,-.5)--(5,-.4);
\draw [thick] [->] (6,-.5)--(6,-.4);
\draw [thick] [->] (7,-.5)--(7,-.4);
\draw [thick] [->] (4,6.5)--(4,6.6);
\draw [thick] [->] (6,6.5)--(6,6.6);
\draw [thick] [->] (1,6.5)--(1,6.4);
\draw [thick] [->] (2,6.5)--(2,6.4);
\draw [thick] [->] (3,6.5)--(3,6.4);
\draw [thick] [->] (5,6.5)--(5,6.4);
\draw [thick] [->] (7,6.5)--(7,6.4);
\draw [thick] [->] (4,4.5)--(4,4.6);
\draw [thick] [->] (6,4.5)--(6,4.6);
\draw [thick] [->] (1,4.5)--(1,4.4);
\draw [thick] [->] (2,4.5)--(2,4.4);
\draw [thick] [->] (3,4.5)--(3,4.4);
\draw [thick] [->] (5,4.5)--(5,4.4);
\draw [thick] [->] (7,4.5)--(7,4.4);
\node at (4,5.5) {$r_2$};
\node at (6,5.5) {$r_1$};
\node at (4,-2) {(b)};
\end{tikzpicture}
\caption{(a) A possible $s$th-row configuration, with $s$ up arrows at
  positions $r_1,\dots,r_s$ (here, $N=7$, $s=2$, $r_1=2$, $r_2=4$);
  (b) The corresponding top and bottom portions resulting from
  splitting the original lattice in correspondence of the $s$th
  row.}
\label{fig-RCP-ZZ}
\end{figure}

An interesting property of the domain-wall six-vertex model, directly
following from its peculiar boundary conditions and from the ice-rule,
is that on the $s$th row (i.e., on the $N$ vertical edges between the
$s$th and the $(s+1)$th horizontal lines, counted from the top, in our
conventions) there are exactly $s$ arrows pointing up.  It is thus
natural to describe  configurations on the $s$th row in terms of the
positions $r_1,\dots,r_s$, of such $s$ up arrows (counted from the
right, and with $1\leq r_1<\dots<r_s\leq N$, see
figure~\ref{fig-RCP-ZZ}). Note that these configurations can
equivalently be described in terms of the complementary set of
integers $\rbar_1,\dots,\rbar_{N-s}$, denoting the position of the
$N-s$ down arrows.

Let us now suppose we are assigned a given $s$th-row configuration on
the $N\times N$, described by the positions $r_1,\dots,r_s$ of the $s$
up arrows, and let us imagine cutting all the vertical edges between
the $s$th and $(s+1)$th horizontal lines of the $N\times N$ lattice,
thus separating it into two smaller lattices, see
figure~\ref{fig-RCP-ZZ}; as a result, we obtain an upper lattice with
$s$ horizontal and $N$ vertical lines, and a lower lattice with $N-s$
horizontal and $N$ vertical lines.  The boundary conditions on these
two lattices are naturally inherited from the depicted procedure,
being of domain wall type on three sides, and with $s$ up arrows at
positions $r_1,\dots,r_s$ on the fourth side. We will denote by
$\psitop{r_1,\dots,r_s}$ and $\psibot{r_1,\dots,r_s}$ the partition
functions of the six-vertex model on the upper and lower sublattices,
respectively.

The partition functions $\psitop{r_1,\dots,r_s}$ and
$\psibot{r_1,\dots,r_s}$ can be viewed as components of off-shell
Bethe states in the context of the algebraic Bethe ansatz.  Besides being
of relevance on their own right, they find application in further
investigations of the limit shape of the model \cite{DR-12}, and in
combinatorics \cite{F-06}.  But our main interest here is in the fact
that $\psitop{r_1,\dots,r_s}$ and $\psibot{r_1,\dots,r_s}$ can be
used, if suitably combined, as building blocks to compute some useful
correlation functions.

Our goal in the present paper is therefore twofold. First, to derive
some convenient representations for the partition functions
$\psitop{r_1,\dots,r_s}$ and $\psibot{r_1,\dots,r_s}$.  Second, to
devise how they should be combined to evaluate more sophisticate
correlation functions.

As for the first goal, we will rely on the QISM \cite{TF-79}, and on
some additional ingredients, developed in \cites{CP-05c,CP-07b}, which
allows to obtain multiple integral representations for these
quantities. Note that some representation for, say,
$\psitop{r_1,\dots,r_s}$, has been known for quite a long time: we are
referring to the `coordinate wavefunction' representation, that
follows from the equivalence of the algebraic and coordinate versions
of the Bethe ansatz \cite{IKR-87}, see appendix \ref{app.wave} for
details. However, for reasons that will become clear below, in
relation to our second goal, such `coordinate wavefunction'
representation is not sufficient for our purposes. In the following we
will work out two additional representations for the components of
the off-shell Bethe states.

The partition functions $\psitop{r_1,\dots,r_s}$ and
$\psibot{r_1,\dots,r_s}$, when considered within QISM, depends on the
spectral parameters $\lambda_1,\dots,\lambda_N$, and
$\nu_1,\dots,\nu_s$, and on $\lambda_1,\dots,\lambda_N$,
$\nu_{s+1},\dots,\nu_N$, respectively.  As functions of the spectral
parameters, these two partition functions can be related to each
other, due to the crossing symmetry of the six-vertex model.

We recall that the crossing symmetry is the symmetry of the Boltzmann
weights under reflection of the vertex (together with the orientation
of the arrows on its edges) with respect to the vertical (or
horizontal) line, with the simultaneous interchange of $a_{\alpha k}$
and $b_{\alpha k}$.  This interchange can be implemented by the
substitution
\begin{equation}\label{crossing}
  \lambda_\alpha \mapsto \pi - \lambda_\alpha,
  \qquad\qquad \nu_k\mapsto  -\nu_k,
\end{equation}
in terms of the spectral parameters.  It is easy to see that the
crossing symmetry of the model implies the relation
\begin{equation}\label{duality1}
\psitop{r_1,\dots,r_s}(\lambda_1,\dots,\lambda_N;\nu_1,\dots,\nu_s)
=\psibot{\rbar_1,\dots,\rbar_{N-s}}
(\pi-\lambda_1,\dots,\pi-\lambda_N;-\nu_1,\dots,-\nu_s).
\end{equation}
By means of this relation, a given representation for, say,
$\psitop{r_1,\dots,r_s}$ immediately leads to a corresponding
representation for $\psibot{r_1,\dots,r_s}$.  However, the dependence
of these two representations on the row configuration is in terms of
two complementary set of integers ($r$'s and $\rbar$'s), while to
combine them conveniently to work out representations of correlation
functions, we need the two representations for
$\psitop{r_1,\dots,r_s}$ and $\psibot{r_1,\dots,r_s}$ to be expressed
both in terms of the same set of integers.

In view of these considerations, in the following we will work out
two distinct and essentially different representations for the two
partition functions, which cannot be obtained one from the other
simply by applying the crossing symmetry relation \eqref{duality1}.
As a matter of fact, these two distinct representations comes out
essentially from two slightly different ways of applying the QISM
machinery. They both appear to differ significantly from the longly
known `coordinate wavefunction' representation.

We turn now to the second main goal of the present paper.  As
mentioned above, $\psitop{r_1,\dots,r_s}$ and
$\psibot{r_1,\dots,r_s}$, if suitably combined, can be used as
building block to construct other correlation functions. As a first,
simple, example, let us mention the row configuration probability,
that is the probability of observing a given configuration of arrows
on a given row of the lattice. More specifically, we denote by
$H_{N,s}^{(r_1,\dots,r_s)}$, the probability of observing on the $s$th
row of the $N\times N$ lattice a configuration with $s$ up arrows at
positions $r_1,\dots,r_s$. This nonlocal correlation function was
introduced in \cite{CP-12}, and, independently, in \cite{FR-10}, in
the context of combinatorics.  It is clear that the row configuration
probability can be written as
\begin{equation}\label{defHNs}
H_N^{(r_1,\dots,r_s)}=\frac{1}{Z_N} \psitop{r_1,\dots,r_s}
\, \psibot{r_1,\dots,r_s},
\end{equation}
thus reconducting its evaluation to that of $\psitop{r_1,\dots,r_s}$
and $\psibot{r_1,\dots,r_s}$.

If we specialize the row configuration probability to the first line,
by setting $s=1$, we obtain the probability of observing the sole
reversed arrow of the first row exactly on the $r$th vertical
edge. This quantity, called one-point boundary correlation function,
was introduced and calculated using QISM in \cite{BPZ-02}.

\begin{figure}
\centering 

\begin{tikzpicture}[scale=.45]
\draw [thick] (0.2,1)--(7.8,1);
\draw [thick] (0.2,2)--(7.8,2);
\draw [thick] (0.2,3)--(7.8,3);
\draw [thick] (0.2,4)--(7.8,4);
\draw [thick] (0.2,5)--(7.8,5);
\draw [thick] (0.2,6)--(7.8,6);
\draw [thick] (0.2,7)--(7.8,7);
\draw [thick] (1,0.2)--(1,7.8);
\draw [thick] (2,0.2)--(2,7.8);
\draw [thick] (3,0.2)--(3,7.8);
\draw [thick] (4,0.2)--(4,7.8);
\draw [thick] (5,0.2)--(5,7.8);
\draw [thick] (6,0.2)--(6,7.8);
\draw [thick] (7,0.2)--(7,7.8);
\draw [thick] [->] (.5,1)--(.4,1);
\draw [thick] [->] (.5,2)--(.4,2);
\draw [thick] [->] (.5,3)--(.4,3);
\draw [thick] [->] (.5,4)--(.4,4);
\draw [thick] [->] (.5,5)--(.4,5);
\draw [thick] [->] (.5,6)--(.4,6);
\draw [thick] [->] (.5,7)--(.4,7);
\draw [thick] [->] (1,7.5)--(1,7.4);
\draw [thick] [->] (2,7.5)--(2,7.4);
\draw [thick] [->] (3,7.5)--(3,7.4);
\draw [thick] [->] (4,7.5)--(4,7.4);
\draw [thick] [->] (5,7.5)--(5,7.4);
\draw [thick] [->] (6,7.5)--(6,7.4);
\draw [thick] [->] (7,7.5)--(7,7.4);
\draw [thick] [->] (7.5,1)--(7.6,1);
\draw [thick] [->] (7.5,2)--(7.6,2);
\draw [thick] [->] (7.5,3)--(7.6,3);
\draw [thick] [->] (7.5,4)--(7.6,4);
\draw [thick] [->] (7.5,5)--(7.6,5);
\draw [thick] [->] (7.5,6)--(7.6,6);
\draw [thick] [->] (7.5,7)--(7.6,7);
\draw [thick] [->] (1,.5)--(1,.6);
\draw [thick] [->] (2,.5)--(2,.6);
\draw [thick] [->] (3,.5)--(3,.6);
\draw [thick] [->] (4,.5)--(4,.6);
\draw [thick] [->] (5,.5)--(5,.6);
\draw [thick] [->] (6,.5)--(6,.6);
\draw [thick] [->] (7,.5)--(7,.6);
\draw [thick] [->] (1.5,7)--(1.4,7);
\draw [thick] [->] (1.5,6)--(1.4,6);
\draw [thick] [->] (2.5,7)--(2.4,7);
\draw [thick] [->] (2.5,6)--(2.4,6);
\draw [thick] [->] (3.5,7)--(3.4,7);
\draw [thick] [->] (3.5,6)--(3.4,6);
\draw [thick] [->] (1,6.5)--(1,6.4);
\draw [thick] [->] (2,6.5)--(2,6.4);
\draw [thick] [->] (3,6.5)--(3,6.4);
\draw [thick] [->] (1,5.5)--(1,5.4);
\draw [thick] [->] (2,5.5)--(2,5.4);
\draw [thick] [->] (3,5.5)--(3,5.4);
\draw [decorate,decoration={brace}]
(3.9,8.1) -- (7.1,8.1) node [midway,yshift=9pt] {$r$};
\draw [decorate,decoration={brace}]
(-.1,5.9) -- (-.1,7.1) node [midway,xshift=-9pt] {$s$};
\node at (4,-2) {(a)};
\end{tikzpicture}
\quad
\begin{tikzpicture}[scale=.45]
\draw [thick] (0.2,1)--(7.8,1);
\draw [thick] (0.2,2)--(7.8,2);
\draw [thick] (0.2,3)--(7.8,3);
\draw [thick] (0.2,4)--(7.8,4);
\draw [thick] (0.2,5)--(7.8,5);
\draw [thick] (0.2,6)--(7.8,6);
\draw [thick] (0.2,7)--(7.8,7);
\draw [thick] (1,0.2)--(1,7.8);
\draw [thick] (2,0.2)--(2,7.8);
\draw [thick] (3,0.2)--(3,7.8);
\draw [thick] (4,0.2)--(4,7.8);
\draw [thick] (5,0.2)--(5,7.8);
\draw [thick] (6,0.2)--(6,7.8);
\draw [thick] (7,0.2)--(7,7.8);
\draw [thick] [->] (.5,1)--(.4,1);
\draw [thick] [->] (.5,2)--(.4,2);
\draw [thick] [->] (.5,3)--(.4,3);
\draw [thick] [->] (.5,4)--(.4,4);
\draw [thick] [->] (.5,5)--(.4,5);
\draw [thick] [->] (.5,6)--(.4,6);
\draw [thick] [->] (.5,7)--(.4,7);
\draw [thick] [->] (1,7.5)--(1,7.4);
\draw [thick] [->] (2,7.5)--(2,7.4);
\draw [thick] [->] (3,7.5)--(3,7.4);
\draw [thick] [->] (4,7.5)--(4,7.4);
\draw [thick] [->] (5,7.5)--(5,7.4);
\draw [thick] [->] (6,7.5)--(6,7.4);
\draw [thick] [->] (7,7.5)--(7,7.4);
\draw [thick] [->] (7.5,1)--(7.6,1);
\draw [thick] [->] (7.5,2)--(7.6,2);
\draw [thick] [->] (7.5,3)--(7.6,3);
\draw [thick] [->] (7.5,4)--(7.6,4);
\draw [thick] [->] (7.5,5)--(7.6,5);
\draw [thick] [->] (7.5,6)--(7.6,6);
\draw [thick] [->] (7.5,7)--(7.6,7);
\draw [thick] [->] (1,.5)--(1,.6);
\draw [thick] [->] (2,.5)--(2,.6);
\draw [thick] [->] (3,.5)--(3,.6);
\draw [thick] [->] (4,.5)--(4,.6);
\draw [thick] [->] (5,.5)--(5,.6);
\draw [thick] [->] (6,.5)--(6,.6);
\draw [thick] [->] (7,.5)--(7,.6);
\draw [thick] [->] (1,5.5)--(1,5.4);
\draw [thick] [->] (2,5.5)--(2,5.4);
\draw [thick] [->] (3,5.5)--(3,5.4);
\node at (4,-2) {(b)};
\end{tikzpicture}
\quad
\begin{tikzpicture}[scale=.45]
\draw [thick] (0.2,1)--(7.8,1);
\draw [thick] (0.2,2)--(7.8,2);
\draw [thick] (0.2,3)--(7.8,3);
\draw [thick] (0.2,4)--(7.8,4);
\draw [thick] (0.2,5)--(7.8,5);
\draw [thick] (0.2,6)--(7.8,6);
\draw [thick] (0.2,7)--(7.8,7);
\draw [thick] (1,0.2)--(1,7.8);
\draw [thick] (2,0.2)--(2,7.8);
\draw [thick] (3,0.2)--(3,7.8);
\draw [thick] (4,0.2)--(4,7.8);
\draw [thick] (5,0.2)--(5,7.8);
\draw [thick] (6,0.2)--(6,7.8);
\draw [thick] (7,0.2)--(7,7.8);
\draw [thick] [->] (.5,1)--(.4,1);
\draw [thick] [->] (.5,2)--(.4,2);
\draw [thick] [->] (.5,3)--(.4,3);
\draw [thick] [->] (.5,4)--(.4,4);
\draw [thick] [->] (.5,5)--(.4,5);
\draw [thick] [->] (.5,6)--(.4,6);
\draw [thick] [->] (.5,7)--(.4,7);
\draw [thick] [->] (1,7.5)--(1,7.4);
\draw [thick] [->] (2,7.5)--(2,7.4);
\draw [thick] [->] (3,7.5)--(3,7.4);
\draw [thick] [->] (4,7.5)--(4,7.4);
\draw [thick] [->] (5,7.5)--(5,7.4);
\draw [thick] [->] (6,7.5)--(6,7.4);
\draw [thick] [->] (7,7.5)--(7,7.4);
\draw [thick] [->] (7.5,1)--(7.6,1);
\draw [thick] [->] (7.5,2)--(7.6,2);
\draw [thick] [->] (7.5,3)--(7.6,3);
\draw [thick] [->] (7.5,4)--(7.6,4);
\draw [thick] [->] (7.5,5)--(7.6,5);
\draw [thick] [->] (7.5,6)--(7.6,6);
\draw [thick] [->] (7.5,7)--(7.6,7);
\draw [thick] [->] (1,.5)--(1,.6);
\draw [thick] [->] (2,.5)--(2,.6);
\draw [thick] [->] (3,.5)--(3,.6);
\draw [thick] [->] (4,.5)--(4,.6);
\draw [thick] [->] (5,.5)--(5,.6);
\draw [thick] [->] (6,.5)--(6,.6);
\draw [thick] [->] (7,.5)--(7,.6);
\draw [dashed] (3.5,8.5)--(3.5,0);
\draw [thick] [->] (4,5.5)--(4,5.6);
\draw [thick] [->] (6,5.5)--(6,5.6);
\node at (4,8.5) {$r_2$};
\node at (6,8.5) {$r_1$};
\draw [decorate,decoration={brace}]
(7.1,-.1) -- (3.9,-.1) node [midway,yshift=-9pt] {$r$};
\node at (4,-2) {(c)};
\end{tikzpicture}

\caption{Emptiness formation probability: (a) Basic definition, as the
  probability of having on all edges within a rectangular region of
  size $(N-r)\times s$ in the top-left corner of the lattice, arrows
  pointing down or left (here is shown the case $s=2$, $r=4$ and
  $N=7$); (b) Equivalent definition, as the probability of having,
  between the $s$th and $(s+1)$th horizontal lines, $N-r$ down arrows
  at positions $N-r+1,\ldots,N$; (c) In terms of the row configuration
  probability, as a sum over $1\leq r_1 <\ldots< r_s\leq r$, where the
  dashed line shows the border which cannot be passed by the positions
  $r_1,\ldots,r_s$ of the up arrows in the summation.}
\label{fig-EFP-NW}
\end{figure}
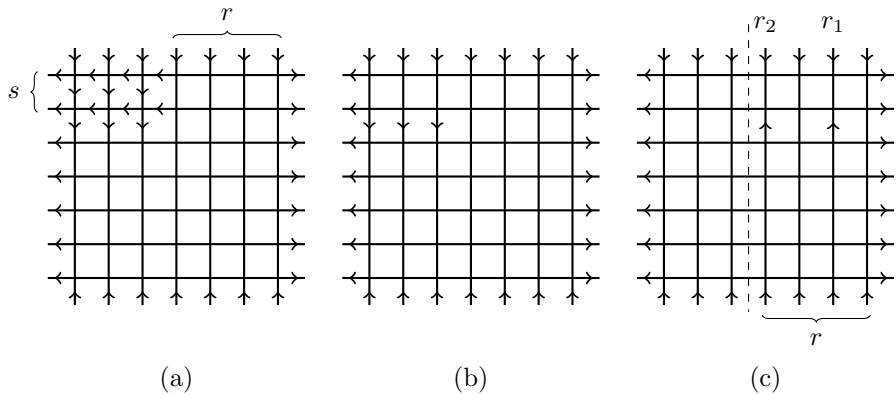

Besides being interesting on its own right, the row configuration
probability can be employed to build useful correlation functions,
such as, for instance, the one-point correlation function, or
polarization. Let us denote by $G_N^{(r,s)}$ the polarization at point
$(r,s)$, that is the probability of observing an up arrow on the $r$th
vertical edge of the $s$th row of the lattice.  Such up arrow may be
any of the $s$ up arrows occuring in the $s$th row.  Let us suppose it
is the $l$th one, $l=1,\dots,s$. We thus have to sum over the
positions of the $l-1$ up arrows to its right, $1\leq
r_1<\dots<r_{l-1}<r$, and over the positions of the $s-l$ up arrows to
its left, $r<r_{l+1}<\dots<r_s\leq N$. The result must then
be summed over all possible values of $l$. We can thus write:
\begin{equation}\label{polarization}
  G_N^{(r,s)}=\sum_{l=1}^s\,\sum_{1\leq r_1<\dots<r_{l-1}<r<r_{l+1}<\dots<r_s\leq N}
  H_N^{(r_1,\dots,r_{l-1},r,r_{l+1},\dots,r_s)}.
\end{equation}
Clearly, one has still to work out some convenient procedure to
perform the multiple sums. As a training ground to tackle such
problem, we will turn our attention towards some slightly simpler,
although nonlocal correlation function, namely, the emptiness
formation probability.

The emptiness formation probability, denoted by $F_N^{(r,s)}$,
describes the probability of having on all edges within a top-left
rectangular region of size $(N-r)\times s$ arrows pointing down or
left, see figure~\ref{fig-EFP-NW}a. It was first introduced in
\cite{CP-07b}, where it was shown to satisfy a recurrence relation in
$r$, $s$ and $N$. Such recurrence relation can be solved, allowing to
build an $s$-fold multiple integral representation.

The domain wall boundary conditions, together with the ice rule, imply
that the emptiness formation probability may be equivalently defined
as the probability of observing the last $N-r$ arrows between the
$s$th and $(s+1)$th horizontal lines to be all pointing down, see
figure~\ref{fig-EFP-NW}b.  It can thus be expressed in terms of the
row configuration probability, as a sum over $1\leq r_1 <\ldots< r_s\leq r$,
see  figure~\ref{fig-EFP-NW}c:
\begin{equation}\label{efp}
  F_N^{(r,s)}=\sum_{r_s=s}^r\dots\sum_{r_2=2}^{r_3-1}\sum_{r_1=1}^{r_2-1}
  H_N^{(r_{1},\dots,r_{s})}.
\end{equation}
The problem of performing the multiple sum appears here to be simpler
with respect to the above mentioned case of polarization, see
\eqref{polarization}. It will be addressed in section 5, where, basing
on the results of \cite{CP-12} and \cite{CCP-19}, we will recover by
different methods the $s$-fold multiple integral representation worked
out in \cite{CP-07b}.

\begin{figure}
\centering

\begin{tikzpicture}[scale=.45]
\draw [thick] (0.2,1)--(7.8,1);
\draw [thick] (0.2,2)--(7.8,2);
\draw [thick] (0.2,3)--(7.8,3);
\draw [thick] (0.2,4)--(7.8,4);
\draw [thick] (0.2,5)--(7.8,5);
\draw [thick] (0.2,6)--(7.8,6);
\draw [thick] (0.2,7)--(7.8,7);
\draw [thick] (1,0.2)--(1,7.8);
\draw [thick] (2,0.2)--(2,7.8);
\draw [thick] (3,0.2)--(3,7.8);
\draw [thick] (4,0.2)--(4,7.8);
\draw [thick] (5,0.2)--(5,7.8);
\draw [thick] (6,0.2)--(6,7.8);
\draw [thick] (7,0.2)--(7,7.8);
\draw [thick] [->] (.5,1)--(.4,1);
\draw [thick] [->] (.5,2)--(.4,2);
\draw [thick] [->] (.5,3)--(.4,3);
\draw [thick] [->] (.5,4)--(.4,4);
\draw [thick] [->] (.5,5)--(.4,5);
\draw [thick] [->] (.5,6)--(.4,6);
\draw [thick] [->] (.5,7)--(.4,7);
\draw [thick] [->] (1,7.5)--(1,7.4);
\draw [thick] [->] (2,7.5)--(2,7.4);
\draw [thick] [->] (3,7.5)--(3,7.4);
\draw [thick] [->] (4,7.5)--(4,7.4);
\draw [thick] [->] (5,7.5)--(5,7.4);
\draw [thick] [->] (6,7.5)--(6,7.4);
\draw [thick] [->] (7,7.5)--(7,7.4);
\draw [thick] [->] (7.5,1)--(7.6,1);
\draw [thick] [->] (7.5,2)--(7.6,2);
\draw [thick] [->] (7.5,3)--(7.6,3);
\draw [thick] [->] (7.5,4)--(7.6,4);
\draw [thick] [->] (7.5,5)--(7.6,5);
\draw [thick] [->] (7.5,6)--(7.6,6);
\draw [thick] [->] (7.5,7)--(7.6,7);
\draw [thick] [->] (1,.5)--(1,.6);
\draw [thick] [->] (2,.5)--(2,.6);
\draw [thick] [->] (3,.5)--(3,.6);
\draw [thick] [->] (4,.5)--(4,.6);
\draw [thick] [->] (5,.5)--(5,.6);
\draw [thick] [->] (6,.5)--(6,.6);
\draw [thick] [->] (7,.5)--(7,.6);
\draw [thick] [->] (7,1.5)--(7,1.6);
\draw [thick] [->] (6,1.5)--(6,1.6);
\draw [thick] [->] (7,2.5)--(7,2.6);
\draw [thick] [->] (6,2.5)--(6,2.6);
\draw [thick] [->] (7,3.5)--(7,3.6);
\draw [thick] [->] (6,3.5)--(6,3.6);
\draw [thick] [->] (6.5,1)--(6.6,1);
\draw [thick] [->] (6.5,2)--(6.6,2);
\draw [thick] [->] (6.5,3)--(6.6,3);
\draw [thick] [->] (5.5,1)--(5.6,1);
\draw [thick] [->] (5.5,2)--(5.6,2);
\draw [thick] [->] (5.5,3)--(5.6,3);
\draw [decorate,decoration={brace}]
(8.1,7.1)--(8.1,3.9) node [midway,xshift=9pt] {$r$};
\draw [decorate,decoration={brace}]
(7.1,-.1)--(5.9,-.1) node [midway,yshift=-9pt] {$s$};
\node at (4,-2) {(a)};
\end{tikzpicture}
\begin{tikzpicture}[scale=.45]
\draw [thick] (0.2,1)--(7.8,1);
\draw [thick] (0.2,2)--(7.8,2);
\draw [thick] (0.2,3)--(7.8,3);
\draw [thick] (0.2,4)--(7.8,4);
\draw [thick] (0.2,5)--(7.8,5);
\draw [thick] (0.2,6)--(7.8,6);
\draw [thick] (0.2,7)--(7.8,7);
\draw [thick] (1,0.2)--(1,7.8);
\draw [thick] (2,0.2)--(2,7.8);
\draw [thick] (3,0.2)--(3,7.8);
\draw [thick] (4,0.2)--(4,7.8);
\draw [thick] (5,0.2)--(5,7.8);
\draw [thick] (6,0.2)--(6,7.8);
\draw [thick] (7,0.2)--(7,7.8);
\draw [thick] [->] (.5,1)--(.4,1);
\draw [thick] [->] (.5,2)--(.4,2);
\draw [thick] [->] (.5,3)--(.4,3);
\draw [thick] [->] (.5,4)--(.4,4);
\draw [thick] [->] (.5,5)--(.4,5);
\draw [thick] [->] (.5,6)--(.4,6);
\draw [thick] [->] (.5,7)--(.4,7);
\draw [thick] [->] (1,7.5)--(1,7.4);
\draw [thick] [->] (2,7.5)--(2,7.4);
\draw [thick] [->] (3,7.5)--(3,7.4);
\draw [thick] [->] (4,7.5)--(4,7.4);
\draw [thick] [->] (5,7.5)--(5,7.4);
\draw [thick] [->] (6,7.5)--(6,7.4);
\draw [thick] [->] (7,7.5)--(7,7.4);
\draw [thick] [->] (7.5,1)--(7.6,1);
\draw [thick] [->] (7.5,2)--(7.6,2);
\draw [thick] [->] (7.5,3)--(7.6,3);
\draw [thick] [->] (7.5,4)--(7.6,4);
\draw [thick] [->] (7.5,5)--(7.6,5);
\draw [thick] [->] (7.5,6)--(7.6,6);
\draw [thick] [->] (7.5,7)--(7.6,7);
\draw [thick] [->] (1,.5)--(1,.6);
\draw [thick] [->] (2,.5)--(2,.6);
\draw [thick] [->] (3,.5)--(3,.6);
\draw [thick] [->] (4,.5)--(4,.6);
\draw [thick] [->] (5,.5)--(5,.6);
\draw [thick] [->] (6,.5)--(6,.6);
\draw [thick] [->] (7,.5)--(7,.6);
\draw [thick] [->] (7,3.5)--(7,3.6);
\draw [thick] [->] (6,3.5)--(6,3.6);
\node at (4,-2) {(b)};
\end{tikzpicture}
\quad
\begin{tikzpicture}[scale=.45]
\draw [thick] (0.2,1)--(7.8,1);
\draw [thick] (0.2,2)--(7.8,2);
\draw [thick] (0.2,3)--(7.8,3);
\draw [thick] (0.2,4)--(7.8,4);
\draw [thick] (0.2,5)--(7.8,5);
\draw [thick] (0.2,6)--(7.8,6);
\draw [thick] (0.2,7)--(7.8,7);
\draw [thick] (1,0.2)--(1,7.8);
\draw [thick] (2,0.2)--(2,7.8);
\draw [thick] (3,0.2)--(3,7.8);
\draw [thick] (4,0.2)--(4,7.8);
\draw [thick] (5,0.2)--(5,7.8);
\draw [thick] (6,0.2)--(6,7.8);
\draw [thick] (7,0.2)--(7,7.8);
\draw [thick] [->] (.5,1)--(.4,1);
\draw [thick] [->] (.5,2)--(.4,2);
\draw [thick] [->] (.5,3)--(.4,3);
\draw [thick] [->] (.5,4)--(.4,4);
\draw [thick] [->] (.5,5)--(.4,5);
\draw [thick] [->] (.5,6)--(.4,6);
\draw [thick] [->] (.5,7)--(.4,7);
\draw [thick] [->] (1,7.5)--(1,7.4);
\draw [thick] [->] (2,7.5)--(2,7.4);
\draw [thick] [->] (3,7.5)--(3,7.4);
\draw [thick] [->] (4,7.5)--(4,7.4);
\draw [thick] [->] (5,7.5)--(5,7.4);
\draw [thick] [->] (6,7.5)--(6,7.4);
\draw [thick] [->] (7,7.5)--(7,7.4);
\draw [thick] [->] (7.5,1)--(7.6,1);
\draw [thick] [->] (7.5,2)--(7.6,2);
\draw [thick] [->] (7.5,3)--(7.6,3);
\draw [thick] [->] (7.5,4)--(7.6,4);
\draw [thick] [->] (7.5,5)--(7.6,5);
\draw [thick] [->] (7.5,6)--(7.6,6);
\draw [thick] [->] (7.5,7)--(7.6,7);
\draw [thick] [->] (1,.5)--(1,.6);
\draw [thick] [->] (2,.5)--(2,.6);
\draw [thick] [->] (3,.5)--(3,.6);
\draw [thick] [->] (4,.5)--(4,.6);
\draw [thick] [->] (5,.5)--(5,.6);
\draw [thick] [->] (6,.5)--(6,.6);
\draw [thick] [->] (7,.5)--(7,.6);
\draw [thick] [->] (7,3.5)--(7,3.6);
\draw [thick] [->] (6,3.5)--(6,3.6);
\draw [dashed] (5.5,0)--(5.5,8.5);
\draw [thick] [->] (2,3.5)--(2,3.6);
\draw [thick] [->] (4,3.5)--(4,3.6);
\node at (7,8.5) {$r_1$};
\node at (6,8.5) {$r_2$};
\node at (4,8.5) {$r_3$};
\node at (2,8.5) {$r_4$};

\draw [decorate,decoration={brace}]
(7.1,-.1) -- (5.9,-.1) node [midway,yshift=-9pt] {$s$};
\node at (4,-2) {(c)};
\end{tikzpicture}

\caption{One more definition of emptiness formation probability: (a)
  The same as in figure~\ref{fig-EFP-NW}a, but reflected with respect
  to the SW-NE diagonal and with all arrows reversed; (b) The same as
  in figure~\ref{fig-EFP-NW}b, but now, between the $r$th and
  $(r+1)$th horizontal lines, among the $r$ up arrows, $s$ of them
  must be at positions $1,\ldots,s$; (c) In terms of row configuration
  probability, as a sum over the positions of the remaining $n=r-s$ up
  arrows $s+1\leq r_{s+1}<\ldots<r_{n+s}\leq N$, where the dashed line
  now shows that all positions at $1,\ldots,s$ are occupied.}
\label{fig-EFP-SE}
\end{figure}
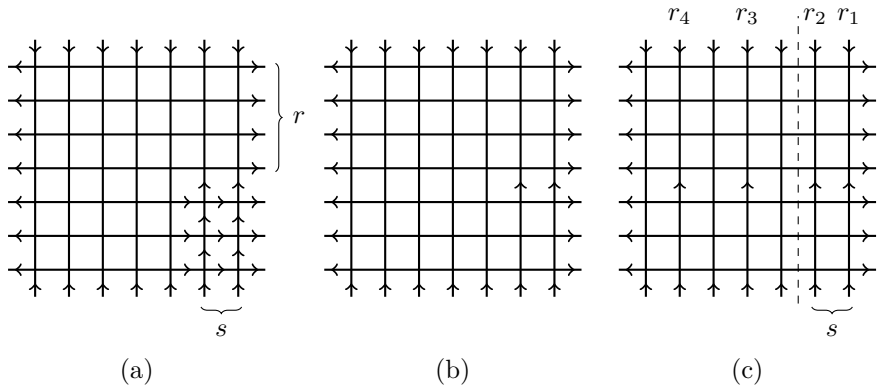

For later use, let us discuss here an alternative way to express the
emptiness formation probability in terms of the row configuration
probability, and of the partition functions $\psitop{r_1,\dots,r_s}$
and $\psibot{r_1,\dots,r_s}$. Recalling that the emptiness formation
probability $F_N^{(r,s)}$ vanishes for $s>r$, it is convenient to
introduce the lattice coordinate $n:=r-s$, with $n=0,1,\dots,N-s$,
giving the distance from the antidiagonal of the lattice. Then, by
definition, $F_N^{(s+n,s)}$ is a weighted sum over all configurations
of the domain-wall $N\times N$ lattice, conditioned to have an
$(N-s-n)\times s$ frozen rectangle in the top-left corner.  Due to
diagonal symmetry, and symmetry under reversal of all arrows, we can
equivalently sum over all configurations conditioned to have a frozen
rectangular region of size $s\times (N-s-n)$ in the bottom right
corner, see figure~\ref{fig-EFP-SE}a.

Let us now focus on the configurations of arrows on the $N$ vertical
edges of the $(s+n)$th row, and denote by $r_1,\dots,r_{s+n}$ the
position of the $s+n$ up arrows. Due to the condition on the
considered configurations, the first $s$ up arrows (counting from the
right) will certainly occur at position $1,\dots,s$, that is $r_j=j$,
$j=1,\dots,s$, see figure~\ref{fig-EFP-SE}b. It can thus be expressed
in terms of the row configuration probability as a sum over the
position of the remaining $n=r-s$ arrows, $s+1\leq
r_{s+1}<\dots<r_{s+n}\leq N$, see figure~\ref{fig-EFP-SE}c:
\begin{equation}\label{efpn}
  F_N^{(s+n,s)}=\sum_{s+1\leq r_{s+1}<\dots<r_{s+n}\leq N}
 H_N^{(1,\dots,s,r_{s+1},\dots,r_{s+n})}.
\end{equation}
This expression will be our starting point in section 6, where,
resorting once more to the technique developed in \cite{CP-12,CCP-19},
we will obtain an $n$-fold (rather than $s$-fold) multiple integral
representation for the emptiness formation probability $F_N^{(r,s)}$.

\subsection{Quantum Inverse Scattering Method formulation}


We now define the main objects of QISM in relation to the model.
First, let us consider vector space $\mathbb{C}^2$ and denote its
basis vectors as the spin-up and spin-down states
\begin{equation}
\ket{\uparrow}=
\begin{pmatrix}
1 \\ 0
\end{pmatrix},\qquad
\ket{\downarrow}=
\begin{pmatrix}
0 \\ 1
\end{pmatrix}.
\end{equation}
To each horizontal and vertical line of the lattice we associate a
copy of the vector space $\mathbb{C}^2$.  We also use the convention
that an upward or right arrow corresponds to a spin-up state while
a downward or left arrow corresponds to a spin-down state.

Next, we introduce the quantum $L$-operator, which can be defined
as a matrix of the Boltzmann weights.  Namely, to each vertex being
intersection of the $\alpha$th vertical line and the $k$th horizontal
line we associate the operator $L_{\alpha,k}(\lambda_\alpha,\nu_k)$
which acts in the direct product of two vector spaces $\mathbb{C}^2$:
in the `horizontal' space $\mathcal{H}_k=\mathbb{C}^2$ (associated
with the $k$th horizontal line) and in the `vertical' space
$\mathcal{V}_\alpha=\mathbb{C}^2$ (associated with the $\alpha$th
vertical line). Referring to the scattering matrix picture, we regard
arrow states on the top and right edges of the vertex as `in' indices
of the $L$-operator while those on the bottom and left edges as `out'
ones. Explicitly, the $L$-operator reads
\begin{equation}\label{Lop}
L_{\alpha,k}(\lambda_\alpha,\nu_k)=
\sin(\lambda_\alpha-\nu_k+\eta\,\tau_\alpha^z \sigma_k^z )
+ \sin 2\eta (\tau_\alpha^{-}\sigma_k^{+}
+\tau_\alpha^{+}\sigma_k^{-}).
\end{equation}
Here $\tau$'s ($\sigma$'s) are Pauli matrices of the corresponding
vertical (horizontal) vector spaces.

Further, we introduce the monodromy matrix, which is an ordered
product of $L$-operators.  We define the monodromy matrix here as a
product of $L$-operators along a vertical line, regarding the
corresponding vertical space $\mathcal{V}_\alpha$ as an `auxiliary'
space, and the tensor product of the $N$ horizontal spaces,  
$\mathcal{H}=\ot_{k=1}^N\mathcal{H}_k$, as the quantum space. In
defining the monodromy matrix it is convenient to think of
$L$-operator as acting in $\mathcal{V}_\alpha\otimes\mathcal{H}$ and,
moreover, writing it as $2$-by-$2$ matrix in $\mathcal{V}_\alpha$,
with the entries being quantum operators (acting in $\mathcal{H}$),
\begin{equation}\label{L-op}
L_{\alpha,k}(\lambda_\alpha,\nu_k)=
\begin{pmatrix}
\sin(\lambda_\alpha-\nu_k+\eta\,\sigma_k^z) &  \sin(2\eta)\,\sigma_k^-\\
\sin(2\eta)\,\sigma_k^+ & \sin(\lambda_\alpha-\nu_k-\eta\,\sigma_k^z)
\end{pmatrix}_{[\mathcal{V}_\alpha]}.
\end{equation}
Here the subscript indicates that this is a matrix in
$\mathcal{V}_\alpha$ and $\sigma_k^{l}$ ($l=+,-,z$) denote quantum
operators in $\mathcal{H}$ acting as Pauli matrices in $\mathcal{H}_k$
and identically elsewhere.

The monodromy matrix is defined as
\begin{align}
T_\alpha(\lambda_\alpha)
& =
L_{\alpha,N}(\lambda_\alpha,\nu_N) \cdots
L_{\alpha,2}(\lambda_\alpha,\nu_2) L_{\alpha,1}(\lambda_\alpha,\nu_1)
\notag\\ &
=\begin{pmatrix}
A(\lambda_\alpha)& B(\lambda_\alpha) \\
C(\lambda_\alpha)& D(\lambda_\alpha)
\end{pmatrix}_{[\mathcal{V}_\alpha]}.
\end{align}
The operators $A(\lambda)=A(\lambda;\nu_1,\dots,\nu_N)$, etc,
act in $\mathcal{H}$.
Operators $A(\lambda)$, $B(\lambda)$, $C(\lambda)$, and $D(\lambda)$,
admit simple graphical
interpretation as vertical lines  of the lattice, with top and bottom
arrows fixed.
Let us introduce `all spins down' and `all spins up' states
\begin{equation}\label{UpDown}
\ket{\Uparrow}=\ot_{k=1}^{N}\ \ket{\uparrow}_k,\qquad
\ket{\Downarrow}=\ot_{k=1}^N\ \ket{\downarrow}_k,
\end{equation}
where $\ket{\uparrow}_k$ and $\ket{\downarrow}_k$ are basis vectors of
$\mathcal{H}_k$. In the case of domain wall boundary conditions each
vertical line corresponds to an operator $B(\lambda_\alpha)$ (where
$\alpha$ is the number of the horizontl line) while vectors
\eqref{UpDown} describe states on the right and left boundaries; the
partition function reads:
\begin{equation}\label{ZBBB}
Z_N= \bra{\Downarrow} B(\lambda_N) \cdots B(\lambda_2)B(\lambda_1)\ket{\Uparrow}.
\end{equation}

To fit the row configuration probability $H_{N,s}^{(r_1,\dots,r_s)}$
into the framework of QISM, we consider the following decomposition of
the monodromy matrix,
\begin{equation}\label{T=T2T1}
T(\lambda)=T\subbot(\lambda) T\subtop(\lambda),
\end{equation}
where $T\subtop(\lambda)$ is defined as a product of the $s$ first
$L$-operators
\begin{equation}\label{T2T1}
T\subtop(\lambda)= L_{\alpha,s}(\lambda,\nu_{s})\cdots
L_{\alpha,1}(\lambda,\nu_1),
\qquad
\end{equation}
and $T\subbot(\lambda)$ as the product of the remaining $N-s$ ones:
\begin{equation}
T\subbot(\lambda)=
L_{\alpha, N}(\lambda,\nu_N) \cdots
L_{\alpha,s+1}(\lambda,\nu_{s+1}).
\end{equation}
We correspondingly decompose the quantum space $\mathcal{H}$ into a
`top' quantum space
$\mathcal{H}^{\mathrm{top}}=\ot_{k=1}^s\mathcal{H}_k$ and a `bottom'
quantum space $\mathcal{H}^{\mathrm{bot}}=\ot_{k=s+1}^N\mathcal{H}_k$,
with
$\mathcal{H}=\mathcal{H}^{\mathrm{top}}\otimes\mathcal{H}^{\mathrm{bot}}$.
Correspondingly, we introduce the operators $A\subtop(\lambda), \dots,
D\subtop(\lambda)$, and $A\subbot(\lambda),\dots, D\subbot(\lambda)$,
as operator valued entries of the corresponding monodromy matrices
$T\subtop(\lambda)$ and $T\subbot(\lambda)$, respectively. Such a
decomposition was originally introduced in the context of the
so-called `two-site model' \cites{IK-84,KBI-93}.

It is useful to consider the corresponding decomposition of the `all
spins up' and `all spin down' vectors. For example, we have
$\ket{\Uparrow}=\ket{\Uparrow\subtop}\ot \ket{\Uparrow\subbot}$,
where, to fit \eqref{T2T1}, we set
\begin{equation}
\ket{\Uparrow\subtop}=\ot_{k=1}^s \ket{\uparrow}_k, \qquad
\ket{\Uparrow\subbot}=\ot_{k=s+1}^N \ket{\uparrow}_k,
\end{equation}
and an analogous decomposition for the `all spins down' vector.  It is
easy to verify that the above defined vectors are eigenvectors of
$A\subtop(\lambda)$, $D\subtop(\lambda)$, and $A\subbot(\lambda)$,
$D\subbot(\lambda)$, respectively.  In particular, we have:
\begin{align}\label{Avac}
A\subbot(\lambda)\ket{\Uparrow\subbot}
&=\prod_{k=s+1}^N a(\lambda,\nu_k)\ket{\Uparrow\subbot},
\\ \label{vacD}
\bra{\Downarrow\subtop}D\subtop(\lambda)
&=\bra{\Downarrow\subtop} \prod_{k=1}^s a(\lambda,\nu_k).
\end{align}

Using the notation introduced above, the partition functions on the
upper, $N\times s$, lattice can be written, in the spirit of
representation \eqref{ZBBB},
\begin{multline}\label{defZ1}
\psitop{r_1,\dots,r_s}=\bra{\Downarrow\subtop}
D\subtop(\lambda_N) \cdots
D\subtop(\lambda_{r_s+1})\,B\subtop(\lambda_{r_s})\,D\subtop(\lambda_{r_s-1})
\\ \times
\cdots
D\subtop(\lambda_{r_1+1})\,B\subtop(\lambda_{r_1})\,D\subtop(\lambda_{r_1-1})\cdots
D\subtop(\lambda_1)
\ket{\Uparrow\subtop}.
\end{multline}
Similarly, for the partition function on the lower, $N\times (N-s)$, lattice
we have
\begin{multline}\label{defZ2}
\psibot{r_1,\dots,r_s}=\bra{\Downarrow\subbot}
B\subbot(\lambda_N) \cdots
B\subbot(\lambda_{r_s+1})\,A\subbot(\lambda_{r_s})\,B\subbot(\lambda_{r_s-1})
\\ \times
\cdots
B\subbot(\lambda_{r_1+1})\,A\subbot(\lambda_{r_1})\,B\subbot(\lambda_{r_1-1})\cdots
B\subbot(\lambda_1)
\ket{\Uparrow\subbot}.
\end{multline}
Formulas \eqref{defZ1} and \eqref{defZ2} are our starting point in
computing $\psibot{r_1,\dots,r_s}$ and $\psitop{r_1,\dots,r_s}$.

\section{The `top' and `bottom' partition functions}

In this section we compute the components of the off-shell Bethe
states (or partition functions) $\psibot{r_1,\dots,r_s}$ and
$\psitop{r_1,\dots,r_s}$ using the technique of the commutation
relations for the entries of the quantum monodromy matrix (the RTT
relation).

\subsection{Fundamental commutation relations}

One of the most basic relations of QISM is the so-called `RLL' relation
\cites{TF-79,KBI-93}, which reads
\begin{equation}
R_{\alpha\alpha'}(\lambda,\lambda')
\big[L_{\alpha k}(\lambda,\nu)\otimes L_{\alpha' k}(\lambda',\nu)\big]=
\big[L_{\alpha k}(\lambda',\nu)\otimes L_{\alpha' k}(\lambda,\nu)\big]
R_{\alpha\alpha'}(\lambda,\lambda').
\end{equation}
Here $R_{\alpha\alpha'}(\lambda,\lambda')$, called the $R$-matrix, is
a matrix acting in the direct product of two auxiliary vector spaces,
$\mathcal{V}_\alpha\otimes\mathcal{V}_{\alpha'}$, and it can be
conveniently represented as a $4$-by-$4$ matrix (we assume that the
first space refers to the $2$-by-$2$ blocks, while the second one to
the entries in the blocks):
\begin{equation}
R_{\alpha\alpha'}(\lambda,\lambda')=
\begin{pmatrix}
f(\lambda',\lambda) & 0 &0 &0 \\
0 & g(\lambda',\lambda) &1 &0 \\
0 &1 & g(\lambda',\lambda) &0 \\
0 &0 &0 & f(\lambda',\lambda)
\end{pmatrix}_{[\mathcal{V}_\alpha\ot\mathcal{V}_{\alpha'}]}.
\end{equation}
Here the functions $f(\lambda',\lambda)$ and $g(\lambda',\lambda)$ are
\begin{equation}\label{fg}
f(\lambda',\lambda)=
\frac{\sin(\lambda-\lambda'+2\eta)}{\sin(\lambda-\lambda')},\qquad
g(\lambda',\lambda)=
\frac{\sin(2\eta)}{\sin(\lambda-\lambda')}.
\end{equation}
It is to be mentioned that here and below
we are mainly following notations and conventions of book \cite{KBI-93}.

The importance of the RLL relation above resides in that it implies
the following relation, which, in turn, can be called RTT relation,
\begin{equation}\label{RTT}
R_{\alpha\alpha'}(\lambda,\lambda')
\big[T_\alpha(\lambda)\otimes T_{\alpha'}(\lambda')\big]=
\big[T_\alpha(\lambda')\otimes T_{\alpha'}(\lambda)\big]
R_{\alpha\alpha'}(\lambda,\lambda').
\end{equation}
This relation contains in total 16 commutation relations, between the
operators $A(\lambda)$, $B(\lambda)$, $C(\lambda)$, and $D(\lambda)$.
In the following we need only some of these commutation relations,
namely
\begin{align} \label{AA}
A(\lambda)\, A(\lambda')&=A(\lambda')\, A(\lambda),
\\  \label{BB}
B(\lambda)\, B(\lambda')&=B(\lambda')\, B(\lambda),
\\  \label{DD}
D(\lambda)\, D(\lambda')&=D(\lambda')\, D(\lambda),
\\  \label{AB2}
A(\lambda)\, B(\lambda')&=
f(\lambda,\lambda')\, B(\lambda')\, A(\lambda)
+g(\lambda',\lambda)\, B(\lambda)\, A(\lambda'),
\\  \label{BA2}
B(\lambda)\, A(\lambda')&=
f(\lambda,\lambda')\, A(\lambda')\, B(\lambda)
+g(\lambda',\lambda)\, A(\lambda)\, B(\lambda'),
\\  \label{DB1}
D(\lambda)\, B(\lambda')&=
f(\lambda',\lambda)\, B(\lambda')\, D(\lambda)
+g(\lambda,\lambda')\, B(\lambda)\, D(\lambda'),
\\  \label{BD1}
B(\lambda)\, D(\lambda')&=
f(\lambda',\lambda)\, D(\lambda')\, B(\lambda)
+g(\lambda,\lambda')\, D(\lambda)\, B(\lambda').
\end{align}
Taking into account relation \eqref{BB} and using relation \eqref{AB2}, one
can obtain, in the usual spirit of the algebraic Bethe ansatz calculation
(see \cites{TF-79,KBI-93}), the relation:
\begin{equation}\label{derivationAB}
A(\lambda_r)\prod_{\beta=1}^{r-1} B(\lambda_\beta)=
\sum_{\alpha=1}^r
\frac{g(\lambda_\alpha,\lambda_r)}{f(\lambda_\alpha,\lambda_r)}
\prod_{\substack{\beta=1\\ \beta\ne\alpha}}^{r} f(\lambda_\alpha,\lambda_\beta)
\prod_{\substack{\beta=1\\ \beta\ne\alpha}}^{r} B(\lambda_\beta) A(\lambda_\alpha).
\end{equation}
Similarly, taking into account \eqref{AA} and using \eqref{BA2}, one obtains
\begin{equation}\label{derivationBA}
B(\lambda_r)\prod_{\beta=1}^{r-1} A(\lambda_\beta)=
\sum_{\alpha=1}^r
\frac{g(\lambda_\alpha,\lambda_r)}{f(\lambda_\alpha,\lambda_r)}
\prod_{\substack{\beta=1\\ \beta\ne\alpha}}^{r} f(\lambda_\alpha,\lambda_\beta)
\prod_{\substack{\beta=1\\ \beta\ne\alpha}}^{r} A(\lambda_\beta)  B(\lambda_\alpha).
\end{equation}
Analogously, relation \eqref{DB1} together with \eqref{BB} give
\begin{equation}\label{derivationDB}
D(\lambda_r)\prod_{\beta=1}^{r-1} B(\lambda_\beta)=
\sum_{\alpha=1}^r
\frac{g(\lambda_r,\lambda_\alpha)}{f(\lambda_r,\lambda_\alpha)}
\prod_{\substack{\beta=1\\ \beta\ne\alpha}}^{r} f(\lambda_\beta,\lambda_\alpha)
\prod_{\substack{\beta=1\\ \beta\ne\alpha}}^{r} B(\lambda_\beta)D(\lambda_\alpha).
\end{equation}
Finally, due to \eqref{DD} and \eqref{BD1}, we have
\begin{equation}\label{derivationBD}
B(\lambda_r)\prod_{\beta=1}^{r-1} D(\lambda_\beta)=
\sum_{\alpha=1}^r
\frac{g(\lambda_r, \lambda_\alpha)}{f(\lambda_r,\lambda_\alpha)}
\prod_{\substack{\beta=1\\ \beta\ne\alpha}}^{r} f(\lambda_\beta,\lambda_\alpha)
\prod_{\substack{\beta=1\\ \beta\ne\alpha}}^{r} D(\lambda_\beta) B(\lambda_\alpha).
\end{equation}
Evidently, decomposition \eqref{T=T2T1} for the monodromy matrices
implies the existence of RTT relations, analogous to
relations \eqref{RTT}, for the `top' and `bottom' quantum spaces.
These, in turn, contains all commutation relations between operators
$A\subtop(\lambda),\dots,D\subtop(\lambda)$, and between operators
$A\subbot(\lambda),\dots,D\subbot(\lambda)$, respectively; below we
use commutation relations \eqref{derivationAB}--\eqref{derivationBD}
for the `top' and `bottom' quantum spaces.

Looking at formulae \eqref{defZ1} and \eqref{defZ2}, we anticipate
that the four relations \eqref{derivationAB}--\eqref{derivationBD} lead
to four different representations, two for $\psitop{r_1,\dots,r_s}$
and two for $\psibot{r_1,\dots,r_s}$.  In particular, the two
resulting representation for, say $\psitop{r_1,\dots,r_s}$, are
essentially different, each one being in turn related through crossing
symmetry to one of the two representations obtained for
$\psibot{r_1,\dots,r_s}$.

\subsection{Application to the `bottom' partition function}\label{sec.Zbot}

We first consider the computation of the lower sublattice partition
function.  Starting from representation \eqref{defZ2}, we can use the
fundamental RTT relations, and, in particular, generalized commutation
relation \eqref{derivationAB} to move all operators
$A\subbot(\lambda)$ to the right, and make them act on
$\ket{\Uparrow\subbot}$, exploiting relation \eqref{Avac}. Indeed,
using $s$ times commutation relation \eqref{derivationAB}, acting on
the right on the vector $\ket{\Uparrow\subbot}$, and multiplying from
the left with the vector $\bra{\Downarrow\subbot}B(\lambda_N)\cdots
B(\lambda_{r_s+1})$, we obtain
\begin{multline}\label{derivation1}
\psibot{r_1,\dots,r_s}=\sum_{\alpha_1=1}^{r_1}
\sum_{\substack{\alpha_2=1\\ \alpha_2\ne\alpha_1}}^{r_2}
\cdots
\sum_{\substack{\alpha_s=1\\ \alpha_s\ne\alpha_1,\,\dots,\alpha_{s-1}}}^{r_s}
\prod_{j=1}^s\prod_{k=s+1}^N  a(\lambda_{\alpha_j},\nu_k)
\prod_{j=1}^s\frac{g(\lambda_{\alpha_j},\lambda_{r_j})}
{f(\lambda_{\alpha_j},\lambda_{r_j})}
\\
\times
\prod_{\substack{\beta_1=1\\ \beta_1 \ne\alpha_1}}^{r_1}
f(\lambda_{\alpha_1},\lambda_{\beta_1})
\prod_{\substack{\beta_2=1\\ \beta_2 \ne\alpha_1,\alpha_2}}^{r_2}
f(\lambda_{\alpha_2},\lambda_{\beta_2})
\ \cdots
\prod_{\substack{\beta_s=1\\ \beta_s \ne\alpha_1,\dots,\alpha_s}}^{r_s}
f(\lambda_{\alpha_s},\lambda_{\beta_s})
\\
\times
Z_{N-s}\left[\lambda_{\alpha_1},\dots,\lambda_{\alpha_s};\nu_1,\dots,\nu_s \right]
\end{multline}
Here
$Z_{N-s}\left[\lambda_{\alpha_1},\dots,\lambda_{\alpha_s};\nu_1,\dots,\nu_s\right]$
denotes the partition function of the domain-wall six-vertex model on
the $(N-s)\times(N-s)$ lattice, with spectral parameters
$\lambda_\alpha$, $\alpha\in\left\{1,2,\dots,N\right\}
\backslash\left\{\alpha_1,\dots,\alpha_s\right\}$ and $\nu_k$,
$k\in\left\{s+1,\dots,N\right\}$; in other words, the square brackets
indicate independence from the enclosed variables, in comparison with
the `original' sets $\lambda_1,\dots,\lambda_N$ and
$\nu_1,\dots,\nu_N$.

To proceed further, it is convenient to  introduce the function
\begin{equation}\label{vrs}
v_{r}(\lambda):=\frac{\prod_{\alpha=r+1}^N d(\lambda_\alpha,\lambda)
\prod_{\alpha=1}^{r-1} e(\lambda_\alpha,\lambda)}{\prod_{k=s+1}^N b(\lambda,\nu_k)},
\end{equation}
where function $d(\lambda,\lambda')$ has been defined in
\eqref{dfunc}, and
\begin{equation}\label{efunc}
e(\lambda,\lambda')=\sin(\lambda-\lambda'+2\eta).
\end{equation}
We now reexpress functions $f(\lambda,\lambda')$ and
$g(\lambda,\lambda')$ appearing in \eqref{derivation1} in terms of
functions $d(\lambda,\lambda')$ and $e(\lambda,\lambda')$, defined in
\eqref{dfunc} and \eqref{efunc}.  We also substitute the 
Izergin-Korepin expression \eqref{ZN} for the partition function
appearing in \eqref{derivation1}.  In this lengthy but standard
computation we arrive at the formula:
\begin{multline}\label{derivation2}
\psibot{r_1,\dots,r_s}=
\frac{
\prod_{\alpha=1}^N\prod_{k=s+1}^N
a(\lambda_{\alpha},\nu_k)b(\lambda_\alpha,\nu_k)}
{\prod_{1\leq\alpha<\beta\leq N}d(\lambda_\beta,\lambda_\alpha)
\prod_{s+1\leq j<k\leq N} d(\nu_j,\nu_k)}
\\
\times\sum_{\alpha_1}^{r_1}
\sum_{\substack{\alpha_2=1\\ \alpha_2\ne\alpha_1}}^{r_2}
\cdots
\sum_{\substack{\alpha_s=1\\ \alpha_s\ne\alpha_1,\dots,\alpha_{s-1}}}^{r_s}
(-1)^{\sum_{j=1}^s(\alpha_j-1)-\sum_{1\leq j<k\leq s} \chi(\alpha_k,\alpha_j)}
\\
\times
\prod_{j=1}^s v_{r_j}(\lambda_{\alpha_j})
\prod_{1\leq j<k\leq s}\frac{1}{e(\lambda_{\alpha_j},\lambda_{\alpha_k})}
\det\caM_{[\alpha_1,\dots,\alpha_s;1,\dots,s]}.
\end{multline}
Here the function $\chi(\alpha,\beta)$  is defined as
\begin{equation}\label{chiab}
  \chi(\alpha,\beta)=\begin{cases}
  0, & \text{if } \alpha<\beta,\\
  1, & \text{otherwise},
  \end{cases}
  \end{equation}
while $\caM_{[\alpha_1,\dots,\alpha_s;1,\dots,s]}$ denotes the
$(N-s)\times(N-s)$ matrix obtained from the matrix $\caM$, see
\eqref{matT}, by removing rows $\alpha_1,\dots,\alpha_s$, and the
first $s$ columns.  Note that, since function $v_{r}(\lambda)$
vanishes for $\lambda=\lambda_{\alpha}$ ($\alpha=r+1,\dots,N$), all
the sums appearing in \eqref{derivation2} can be extended up to the
value $N$. Finally, note that the functions $e(\lambda,\mu)$ appearing
in the denominator in last line of \eqref{derivation2} are exactly
compensated by corresponding functions in the numerators, that are
hidden in the definition of functions $v_r(\lambda)$. As a
consequence, each term of the sum remain regular even in the limit
where two $\lambda$'s differ exactly by $2\eta$.

It is worth to comment that in computing $\psibot{r_1,\dots,r_s}$ here
we could have proceeded differently. Namely, starting from
representation \eqref{defZ2}, we could have chosen to use commutation
relation \eqref{derivationBA} to move all operators
$A\subbot(\lambda)$ to the left, and make them act on
$\ket{\Uparrow\subbot}$. As a result, $\psibot{r_1,\dots,r_s}$ would
have been expressed as an $(N-s)$-fold sum of minors of order $N-s$ of
the matrix $\caM$, see \eqref{matT}. In this way we would have arrived
at an essentially different representation, in comparison with
\eqref{derivation2}. As it becomes clear below, it is the combination
of these two complementary representations, one for
$\psibot{r_1,\dots,r_s}$ and another for $\psitop{r_1,\dots,r_s}$,
that may lead to useful representations for the row configuration probability
and other correlation functions therefrom.

\subsection{Application to the `top' partition function}

Let us turn to the partition function on the upper sublattice.  Having
in mind representation \eqref{defZ1} for $\psitop{r_1,\dots,r_s}$, we
can use the fundamental RTT relations to move all
$B\subtop(\lambda)$'s on one side, and all $D\subtop(\lambda)$'s on
the other.  As already outlined on the example of
$\psibot{r_1,\dots,r_s}$, one can implement this procedure in two
different ways.  The first possibility is to use commutation relation
\eqref{derivationDB} to commute each of the $D\subtop(\lambda)$'s to
the right, through $B\subtop(\lambda)$'s.  The resulting expression
appears to be dual (under crossing symmetry) to the one computed in
section 3.2. We report it in appendix \ref{app.dual} for the sake of
completeness.

Here we exploit the second possibility, namely, we commute the 
$B\subtop(\lambda)$'s to the right through the $D\subtop(\lambda)$'s. More
specifically, we can use commutation relation \eqref{BD1} to move all
$D\subtop(\lambda)$'s to the left, and make them act on
$\bra{\Downarrow\subtop}$, exploiting relation \eqref{vacD}.  Using
$s$ times commutation relation \eqref{derivationBD}, acting on the
right on the vector $\ket{\Uparrow\subtop}$, multiplying from the left
with the vector $\bra{\Downarrow\subtop}D(\lambda_N)\cdots
D(\lambda_{r_s+1})$, we obtain
\begin{multline}\label{topderivation1bis}
\psitop{r_1,\dots,r_s}=
\sum_{\alpha_1=1}^{r_1}
\sum_{\substack{\alpha_2=1\\ \alpha_2\ne\alpha_1}}^{r_2}
\cdots
\sum_{\substack{\alpha_s=1\\ \alpha_s\ne\alpha_1,\,\dots,\alpha_{s-1}}}^{r_s}
\
\prod_{\substack{\beta=1\\ \beta\ne \alpha_1,\dots,\alpha_s}}^N\prod_{k=1}^{s}
a(\lambda_{\beta},\nu_k)
\prod_{j=1}^s\frac{g(\lambda_{r_j},\lambda_{\alpha_j})}{f(\lambda_{r_j},\lambda_{\alpha_j})}
\\
\times
\prod_{\substack{\beta_1=1\\ \beta_1\ne\alpha_1}}^{r_1}
f(\lambda_{\beta_1},\lambda_{\alpha_1})
\prod_{\substack{\beta_2=1\\ \beta_2\ne\alpha_1,\alpha_2}}^{r_2}
f(\lambda_{\beta_2},\lambda_{\alpha_2})\cdots
\prod_{\substack{\beta_s=1\\ \beta_s\ne\alpha_1,\dots,\alpha_s}}^{r_s}
f(\lambda_{\beta_s},\lambda_{\alpha_s})
\\
\times
Z_{s}\left(\lambda_{\alpha_1},\dots,\lambda_{\alpha_s};\nu_1,\dots,\nu_s \right).
\end{multline}
Here
$Z_{s}\left(\lambda_{\alpha_1},\dots,\lambda_{\alpha_s};\nu_1,\dots,\nu_s
\right)$ denotes the partition function of the inhomogeneous model on
the $s \times s$ lattice, with the indicated spectral parameters.
Substituting the Izergin-Korepin partition function, one can further
rewrite \eqref{topderivation1bis} as a multiple sum involving $s\times
s$ determinant; for our purposes below the above representation
appears to be sufficient.

As a side comment, it is worth to mention that for $s=1$ the formula
\eqref{topderivation1bis} reduces to a sum of $r_1$ terms. On the
other hand, evaluating the weight of the sole possible configuration,
one can find that
\begin{equation}\label{ztops=1}
  \psitop{r_1}=\prod_{\alpha=r_1+1}^N a(\lambda_\alpha,\nu_1)
  \  c \prod_{\alpha=1}^{r_1-1}
b(\lambda_\alpha,\nu_1).
 \end{equation}
Apparently, the equivalence of the two expressions is due to certain
identity; a direct proof of the relevant identity can be found in
\cite{G-83} (see also \cite{BPZ-02}). A natural question is whether
such identity generalizes to values of $s>1$. The answer appears to be
positive.  We report such remarkable identity and discuss some
particular cases in appendix~\ref{app.id}.

To conclude this section, we wish to emphasize that representations
\eqref{derivation1}, or equivalently \eqref{derivation2}, for
$\psibot{r_1,\dots,r_s}$, and \eqref{topderivation1bis} for
$\psitop{r_1,\dots,r_s}$, depend on the row configuration through
$r_1,\dots,r_s$ denoting the positions of the up arrows.  We recall
that these representations have been obtained by repeated use of
\eqref{derivationAB} and \eqref{derivationBD}, respectively. If
instead we had worked out analogous derivations starting from
relations \eqref{derivationDB} and \eqref{derivationBA}, we would have
obtained two different representations, depending on the row
configuration through the complementary set of integers
$\rbar_1,\dots,\rbar_{N-s}$ denoting the position of the down arrows
(see appendix \ref{app.dual}, formula \eqref{derivation22} for
$\psitop{r_1,\dots,r_s}$).  It is clear that the two additional
representations are simply related to \eqref{topderivation1bis} and
\eqref{derivation2} by the crossing symmetry transformation, see
\eqref{duality1}.

Finally, we stress once more that  such representations are
all essentially different from the so-called `coordinate wavefunction'
representation, that follows  from the equivalence of the algebraic and
coordinate Bethe ansatz \cite{IKR-87}. In particular, referring to
$\psitop{r_1,\dots,r_s}$ for definiteness, representations
\eqref{topderivation1bis}, and \eqref{derivation22} are both 
different from \eqref{oldrep}.

\section{Integral representations for the `top' and `bottom' partition functions}

In this section we derive representations for $\psibot{r_1,\dots,r_s}$ 
and $\psitop{r_1,\dots,r_s}$ in terms of $s$-fold contour integrals. 

\subsection{Orthogonal polynomial representation for
  the `bottom' partition function}\label{Sect41}

Let us first consider $\psibot{r_1,\dots,r_s}$. To start with, we
evaluate the homogeneous limit for expression \eqref{derivation2}. We
resort to the procedure successfully used in \cites{CP-05c,CP-07b}. It
is based on the observation that the multiple sum in
\eqref{derivation2} reminds the Laplace expansion of some determinant,
since the minor appearing in the last line depends on the summed
indices only through their absence. Thus the first step is to rewrite
representation \eqref{derivation2} in a determinant form. For this
purpose we set
\begin{equation}
\lambda_\alpha=\lambda+\xi_\alpha, \qquad\alpha=1,\dots,N
\end{equation}
where the $\xi$'s will be sent to zero in the limit (as well as the
$\nu$'s). Keeping the $\xi$'s nonzero (and different from each other),
and using the fact that for a function $f(x)$, regular near
$x=\lambda$, the relation $\exp(\xi\partial_\eps)
f(\lambda+\eps)|_{\eps=0}= f(\lambda+\xi)$ is valid, we can bring
\eqref{derivation2} to the form
\begin{multline}\label{Z2det}
\psibot{r_1,\dots,r_s}=
\frac{
\prod_{\alpha=1}^N\prod_{k=s+1}^N
a(\lambda_{\alpha},\nu_k)b(\lambda_\alpha,\nu_k)}
{\prod_{1\leq\alpha<\beta\leq N}d(\lambda_\beta,\lambda_\alpha)
\prod_{s+1\leq j<k\leq N} d(\nu_j,\nu_k)}
\\ \times
\begin{vmatrix}
\exp(\xi_1\partial_{\eps_1}) & \dots & \exp(\xi_1\partial_{\eps_s})
& \varphi(\lambda_1,\nu_{s+1}) & \dots & \varphi(\lambda_1,\nu_{N})
\\
\hdotsfor{6}
\\
\exp(\xi_N\partial_{\eps_1}) & \dots & \exp(\xi_N\partial_{\eps_s})
& \varphi(\lambda_N,\nu_{s+1}) & \dots & \varphi(\lambda_N,\nu_{N})
\end{vmatrix}
\\ \times\prod_{j=1}^s v_{r_j}(\lambda+\eps_j)
\prod_{1\leq j<k\leq s}^{}
\frac{1}
{e(\lambda+\eps_j,\lambda+\eps_k)}\Bigg|_{\eps_1,\ldots,\eps_s=0}.
\end{multline}
It is to be emphasized that this expression is still for the
inhomogeneous model; it represents an
equivalent way of writing the multiple sum in \eqref{derivation2}.

We can now perform the homogeneous limit along the lines of
\cite{ICK-92}.  Specifically, we send $\xi_1,\dots,\xi_N$ and
$\nu_1,\dots,\nu_N$ to zero. The procedure is explained in full detail
in \cite{CP-07b}.  Factoring out the partition function of the entire
lattice, see \eqref{ZNhom}, we obtain
\begin{multline}\label{homZ2}
\psibot{r_1,\dots,r_s}=
\frac{Z_N\prod_{j=1}^s(N-j)!}{\left(ab\right)^{Ns}\det \caN}
\\ \times
\begin{vmatrix}
\varphi(\lambda)& \dots & \partial_\lambda^{N-s-1}\varphi(\lambda)& 1& \dots &  1
\\
\partial_\lambda\varphi(\lambda) & \dots & \partial_\lambda^{N-s}\varphi(\lambda)
& \partial_{\eps_1} & \dots & \partial_{\eps_s}
\\ \hdotsfor{6} \\
\partial_\lambda^{N-1}\varphi(\lambda) & \dots &
\partial_\lambda^{2N-s-2}\varphi(\lambda)
& \partial_{\eps_1}^{N-1} & \dots & \partial_{\eps_s}^{N-1}
\end{vmatrix}
\\ \times \prod_{j=1}^{s}
\frac{(\sin\eps_j)^{N-r_j}[\sin(\eps_j-2\eta)]^{r_j-1}}
     {[\sin(\eps_j+\lambda-\eta)]^{N-s}}
    \\ \times
     \prod_{1\leq j<k \leq s}^{}
\frac{1}{\sin(\eps_j-\eps_k+2\eta)} \Bigg|_{\eps_1,\ldots,\eps_s=0},
\end{multline}
where, in writing the determinant, we  have changed the order of
columns with respect to \eqref{Z2det}. Here and below, when
considering the homogeneous model we use the short  notation for the
weights, $a=a(\lambda,0)$, $b=b(\lambda,0)$, and for the
partition function, $Z_N=Z_N(\lambda,\dots,\lambda;0,\dots,0)$.

In order to rewrite \eqref{homZ2} in integral form, we first transform
the $N\times N$ determinant into some more convenient and smaller
$s\times s$ determinant, given in terms of a set of orthogonal
polynomials.  These polynomials naturally emerge when the determinant
of matrix $\caN$ entering the homogeneous partition function
\eqref{ZNhom} is interpreted as a Gram determinant associated to
certain integral measure.

The derivation of the $s\times s$ determinant representation from
\eqref{homZ2} is based on the following facts. Let
$\{P_n(x)\}_{n=0}^\infty$ be a set of orthogonal polynomials,
\begin{equation}\label{ortho}
\int   P_{n}(x) P_{m}(x) \mu(x)\,\rmd x
=h_{n} \delta_{nm} ,
\end{equation}
where the integration domain is assumed over the real axis. The weight
$\mu(x)$ is real nonnegative and we choose $h_n$'s such that
$P_n(x)=x^n + \dots$, i.e., that the leading coefficient of $P_n(x)$
is equal to one. Let $c_n$ denote the $n$th moment of the weight
$\mu(x)$,
\begin{equation}
c_n =\int x^n \mu(x)\, \rmd x
\qquad n=0,1,\ldots.
\end{equation}
The orthogonality condition \eqref{ortho} and standard properties of
determinants allow us to prove that
\begin{equation}
\begin{vmatrix}
c_0&c_1& \dots &c_{n-1} \\
c_1& c_2&  \dots & c_{n}\\
\hdotsfor{4}  \\
c_{n-1} & c_{n} & \dots & c_{2n-2}
\end{vmatrix}
= h_0 h_1\cdots h_{n-1}.
\end{equation}
More generally (see, e.g., \cite{S-75}), for $s=1,\dots,N$, the
following formula is valid
\begin{multline}\label{detdet}
\begin{vmatrix}
c_0&c_1& \dots &c_{N-s-1}  & 1 & 1 & \dots & 1 \\
c_1&c_2& \dots &c_{N-s} & x_1 & x_2 &\dots &x_s \\
\hdotsfor{8}  \\
c_{N-1}& c_{N} &\dots & c_{2N-s-2} & x_1^{N-1} & x_2^{N-1} &\dots &x_s^{N-1}
\end{vmatrix}
\\
=h_0 \cdots h_{N-s-1}
\begin{vmatrix}
P_{N-s}(x_1) & \dots & P_{N-s}(x_s)\\
\hdotsfor{3} \\
P_{N-1}(x_1) & \dots & P_{N-1}(x_s)
\end{vmatrix}.
\end{multline}
In our case, $c_n=\partial^n_\lambda\varphi(\lambda)$, and the
integration measure $\mu(x)\rmd x$ is given by the Laplace transform
of the function $\varphi(\lambda)$; for explicit expressions, see
\cite{Zj-00}.

Following \cites{CP-06,CP-05c,CP-07b}, we denote
\begin{equation}\label{Knx}
K_n(x)= \frac{n!\,\varphi^{n+1}}{h_n}\; P_n(x),
\end{equation}
where  $\varphi:=\varphi(\lambda)$, and $h_n$ is given in \eqref{ortho}.
We also introduce the functions
\begin{equation}\label{omegas}
\omega(\epsilon):=\frac{\sin(\lambda+\eta)}{\sin(\lambda-\eta)}\,
\frac{\sin\eps}{\sin(\eps-2\eta)},\qquad
\tilde\omega(\epsilon):=\frac{\sin(\lambda-\eta)}{\sin(\lambda+\eta)}\,
\frac{\sin\eps}{\sin(\eps+2\eta)}.
\end{equation}
These are functions of $\eps$, with $\lambda$ and $\eta$ regarded as
parameters.

Noting that
\begin{equation}
\frac{\sin(\eps_1+\lambda+\eta)\sin(\eps_2+\lambda-\eta)}
{\sin(\eps_1-\eps_2+2\eta)}
=\frac{1}{\varphi}
\frac{[1-\tilde\omega(\eps_1)][\omega(\eps_2)-1]}
     {\tilde\omega(\eps_1)\omega(\eps_2)-1},
\end{equation}
we have, in virtue of \eqref{detdet}, the following orthogonal
polynomials representation:
\begin{multline}\label{orthZ2}
\psibot{r_1,\dots,r_s}=
\frac{Z_N}{a^{\frac{s(2N-s+1)}{2}}
b^{\frac{s(s-3)}{2}}c^s} \prod_{j=1}^{s}\left(\frac{a}{b}\right)^{r_j}
\\ \times
\begin{vmatrix}
K_{N-s}(\partial_{\eps_1}) &\hdots & K_{N-s}(\partial_{\eps_s})\\
\hdotsfor{3} \\
K_{N-1}(\partial_{\eps_1}) &\hdots & K_{N-1}(\partial_{\eps_s})
\end{vmatrix}
\prod_{j=1}^{s}\left\{
\frac{[\omega(\eps_j)]^{N-r_j-s+j}
  [\tilde\omega(\eps_j)]^{s-j}}{[\omega(\eps_j)-1]^{N-s}}
 \right\}
\\ \times
\prod_{1\leq j<k \leq s}^{} \frac{1}
{\tilde\omega(\eps_j)\omega(\eps_k)-1}
\Bigg|_{\eps_1,\ldots,\eps_s=0}.
\end{multline}
This representation is valid for arbitrary values of the parameters of
the model, independently of the regime.

\subsection{Integral representation for  the `bottom' partition function}

Our aim now is to rewrite representation \eqref{orthZ2} as a
multiple integral.  This can be done using the procedure provided in
\cite{CP-07b}, where it was worked out on the example of the emptiness
formation probability.

A special role below is played by the one-point boundary correlation
function, denoted $H_N^{(r)}$, which is exactly the $s$th-row
configuration probability $H_{N,s}^{(r_1,\dots,r_s)}$ in the special
case of $s=1$.  In this case the partition function $\psitop{r}$ can
be easily computed with the result $\psitop{r}=a^{N-r}b^{r-1} c$,
while $\psibot{r}$ can be found from \eqref{orthZ2}, that yields
\begin{equation}\label{H_N}
  H_N^{(r)}=
  K_{N-1}(\partial_\eps)\,\frac{[\omega(\eps)]^{N-r}}{[\omega(\eps)-1]^{N-1}}
  \bigg|_{\epsilon=0}.
\end{equation}
The whole procedure of transforming representation \eqref{orthZ2} into
a multiple integral representation is based on the following key
identity (see \cite{CP-07b} for a proof):
\begin{equation}\label{claim}
K_{N-1}(\partial_\eps)\, f(\omega(\eps))\Big|_{\eps=0}=
\frac{1}{2\pi \rmi}\oint_{C_0}^{} \frac{(z-1)^{N-1}}{z^N} h_N(z) f(z)\, \rmd z.
\end{equation}
Here, $f(z)$ is an arbitrary function regular at the origin, $C_0$ is a
small simple closed counterclockwise contour around the point $z=0$,
and $h_N(z)$ (not to be confused with $h_n$ in \eqref{ortho}) is the
generating function of the one-point boundary correlation function
\eqref{H_N},
\begin{equation}\label{hNz}
h_N(z)=\sum_{r=1}^{N} H_N^{(r)}z^{r-1}.
\end{equation}
Clearly, $h_N(0)=H_N^{1}=a^{2(N-1)}cZ_{_N-1}$, and $h_N(1)=1$.

Further, we introduce functions $h_{N,s}(z_1,\dots,z_s)$, where the
second subscript, $s=1,\dots,N$, refers to the number of
arguments. These functions are defined as
\begin{multline}\label{hNs}
h_{N,s}(z_1,\dots,z_s) =
\prod_{1\leq j<k \leq s}^{} (z_k-z_j)^{-1}
\\ \times
\begin{vmatrix}
z_1^{s-1} h_{N-s+1}(z_1)&  \hdots & z_s^{s-1} h_{N-s+1}(z_s) \\
z_1^{s-2}(z_1-1) h_{N-s+2}(z_1)& \hdots
&  z_s^{s-2}(z_s-1) h_{N-s+2}(z_s) \\
\hdotsfor{3} \\
(z_1-1)^{s-1} h_{N}(z_1)& \hdots
& (z_s-1)^{s-1} h_{N}(z_s)
\end{vmatrix}.
\end{multline}
The functions $h_{N,s}(z_1,\dots,z_s)$ are symmetric polynomials of
degree $N-1$ in each of their variables. Due to the structure of
\eqref{hNs}, it is easy to check the relations
\begin{equation}\label{hns1}
h_{N,s}(z_1,\dots,z_{s-1},1)=h_{N,s-1}(z_1,\dots,z_{s-1})
\end{equation}
and
\begin{equation}\label{hns0}
h_{N,s}(z_1,\dots,z_{s-1},0)=h_N(0)h_{N-1,s-1}(z_1,\dots,z_{s-1}),
\end{equation}
which will play a crucial role below. The functions
$h_{N,s}(z_1,\dots,z_s)$ can be viewed as the multi-variable
generalizations of $h_N(z)$ and turn out to be alternative
representations (with respect to the Izergin-Korepin partition
function) for the partially inhomogeneous partition functions
\cite{CP-07b}.

To proceed, in representation \eqref{orthZ2} it is convenient to
express $\tilde\omega(\eps)$ in terms of $\omega(\eps)$, by means of
the identity
\begin{equation}\label{tildez}
\tilde \omega(\epsilon) = \frac{t^2\omega(\eps)}{2\Delta t \omega(\eps) -1},
\end{equation}
where we have used the parametrization 
\begin{equation}\label{Deltat}
\Delta:=\frac{a^2+b^2-c^2}{2 a b},\qquad t:=\frac{b}{a}.
\end{equation}

Next, resorting to identity \eqref{claim}, we obtain the multiple
integral representation
\begin{multline}\label{MIRZ2}
\psibot{r_1,\dots,r_s}=
Z_N\frac{\prod_{j=1}^{s}t^{j-r_j}}{a^{s(N-1)}c^s}
\oint_{C_0}^{} \cdots \oint_{C_0}^{}
\prod_{j=1}^s\frac{1}{z^{r_j}_j}
\prod_{1\leq j <k \leq s}\frac{z_k-z_j}{t^2 z_j z_k -2\Delta t z_j+1}
\\ \times
h_{N,s}(z_1,\dots,z_s)
\, \frac{\rmd^sz}{(2\pi \rmi)^s}.  
\end{multline}
This representation is the final formula for $\psibot{r_1,\dots,r_s}$.
It is valid for arbitrary values of the parameters of the model,
independently of the regime.

Note that the expression \eqref{MIRZ2} also implies, through the
crossing symmetry transformation, an analogous $(N-s)$-fold integral
representation for $\psitop{r_1,\dots,r_s}$, which depends on the row
configuration through the positions $\rbar_1,\dots,\rbar_{N-s}$ of the
$N-s$ down arrows, for details, see appendix  
\ref{app.dual}, formula  \eqref{MIRZ1}.

\subsection{Integral representation for the `top' partition function}

Let us now turn to the `top' partition function,
$\psitop{r_1,\dots,r_s}$. Before proceeding, it is worth to mention
that a multiple integral representation for such quantity has already
been worked out in \cite{CP-12}, basing on the well-known `coordinate
wavefunction' representation \eqref{oldrep}, that follows from the
equivalence of the algebraic and coordinate Bethe ansatz
\cite{IKR-87}. It reads
\begin{multline}\label{oldrep_mir}
  \psitop{r_1,\dots,r_s}
  = c^s a^{s(N-1)}
     \prod_{j=1}^s t^{r_j-j} \oint_{C_1}^{} \cdots \oint_{C_1}^{}
     \prod_{j=1}^s \frac{w_j^{r_j-1}}{(w_j-1)^s}
     \\
     \times
     \prod_{1\leq j<k\leq s}
     \left[(w_j-w_k)(t^2 w_j w_k -2\Delta t w_j +1)\right]
  \frac{\rmd^s w}{(2\pi\rmi)^s},
\end{multline}
see appendix \ref{app.wave} for a derivation.

We will now derive another, significantly different multiple integral
representation, which will play a crucial role in the following.  Let
us turn back to representation \eqref{topderivation1bis} for
$\psitop{r_1,\dots,r_s}$ and note that the multiple sum therein can be
interpreted as the sum of residues of some function in the $s$-fold
complex plane. Specifically, we can use the following identity
\begin{multline}\label{eqapp6}
\sum_{\alpha_1=1}^{r_1}
\sum_{\substack{\alpha_2=1\\ \alpha_2\ne\alpha_1}}^{r_2}
\cdots
\sum_{\substack{\alpha_s=1\\ \alpha_s\ne\alpha_1,\,\dots,\alpha_{s-1}}}^{r_s}
F(\lambda_{\alpha_1},\dots,\lambda_{\alpha_s})
\\
\times
\prod_{\substack{\beta_1=1\\ \beta_1 \ne\alpha_1}}^{r_1}
\frac{1}{d(\lambda_{\alpha_1},\lambda_{\beta_1})}
\prod_{\substack{\beta_2=1\\ \beta_2 \ne\alpha_1,\alpha_2}}^{r_2}
\frac{1}{d(\lambda_{\alpha_2},\lambda_{\beta_2})} \cdots
\prod_{\substack{\beta_s=1\\ \beta_s \ne\alpha_1,\dots,\alpha_s}}^{r_s}
\frac{1}{d(\lambda_{\alpha_s},\lambda_{\beta_s})}
\\
=\oint_\Clambdas\cdots \oint_\Clambdas
\frac{\prod_{1\leq j< k \leq s} d(\zeta_k,\zeta_j)
}{
\prod_{\beta_1=1}^{r_1}d(\zeta_1,\lambda_{\beta_1})
\cdots\prod_{\beta_s=1}^{r_s}d(\zeta_s,\lambda_{\beta_s})
}
\\ \times
F(\zeta_1,\dots,\zeta_s)
\,\frac{\rmd^s \zeta}{(2\pi\rmi)^s},
\end{multline}
where $\Clambdas:=C_{\lambda_1}\cup\dots\cup C_{\lambda_{r_s}}$ is a
simple closed counterclockwise contour in the complex plane of the
integration variable, enclosing points
$\lambda_1$,$\dots$,$\lambda_{r_s}$ and no other singularity of the
integrand. The function $F(\zeta_1,\dots,\zeta_s)$ is a generic
analytic function of its variables, regular in each variable within
the region delimited by $\Clambdas$.

We now reexpress functions $f(\lambda_\alpha,\lambda_\beta)$ and
$g(\lambda_\alpha,\lambda_\beta)$ appearing in representation
\eqref{topderivation1bis} in terms of functions
$d(\lambda_\alpha,\lambda_\beta)$ and
$e(\lambda_\alpha,\lambda_\beta)$, see \eqref{dfunc} and
\eqref{efunc}. Comparing the resulting expression with the left hand
side of identity \eqref{eqapp6}, we  set
\begin{multline}\label{functionF}
F(\lambda_{\alpha_1},\dots,\lambda_{\alpha_s})=\prod_{\beta=1}^N\prod_{k=1}^s
a(\lambda_\beta,\nu_k)\prod_{j,k=1}^s\frac{1}{a(\lambda_{\alpha_j},\nu_k)}
\\
\times
\prod_{\beta_1=1}^{r_1-1} e(\lambda_{\alpha_1},\lambda_{\beta_1})\cdots
\prod_{\beta_s=1}^{r_s-1} e(\lambda_{\alpha_s},\lambda_{\beta_s})
\prod_{1\leq j<k \leq s} \frac{1}{e(\lambda_{\alpha_k},\lambda_{\alpha_j})}
\\
\times
Z_s(\lambda_{\alpha_1},\dots,\lambda_{\alpha_s};\nu_1,\dots,\nu_s).
\end{multline}
As a result, we obtain the following multiple integral representation:
\begin{multline}\label{topderivation2}
\psitop{r_1,\dots,r_s}=
\prod_{\beta=1}^N\prod_{k=1}^s a(\lambda_{\beta},\nu_k)
\\ \times
\oint_\Clambdas\cdots \oint_\Clambdas
\frac{\prod_{\beta_1=1}^{r_1-1}e(\zeta_1,\lambda_{\beta_1})
\cdots\prod_{\beta_s=1}^{r_s-1}e(\zeta_s,\lambda_{\beta_s})
}{
\prod_{\beta_1=1}^{r_1}d(\zeta_1,\lambda_{\beta_1})
\cdots\prod_{\beta_s=1}^{r_s}d(\zeta_s,\lambda_{\beta_s})
}
\\ \times
\prod_{1\leq j<k\leq s}\frac{d(\zeta_k,\zeta_j)}{e(\zeta_k,\zeta_j)}\,
\frac{Z_{s}\left(\zeta_1,\dots,\zeta_s;\nu_1,\dots,\nu_s \right)}
{\prod_{j=1}^s\prod_{k=1}^s a(\zeta_j,\nu_k)}
\,\frac{\rmd^s \zeta}{(2\pi \rmi)^s}.
\end{multline}
It is worth to recall that, besides the already mentioned poles at
$\lambda_1,\dots,\lambda_{r_s}$, the only other poles of the integrand
within the strip of width $\pi$ originate from the functions
$a(\zeta_j,\nu_k)$ appearing in the denominator. As for $Z_{s}$, it is
a regular function of its spectral parameters. Concerning the
functions $e(\zeta_k,\zeta_j)$ in the denominator, the poles they give
rise to are only apparent. Indeed, focussing \eqref{functionF}, a
careful comparison with \eqref{topderivation1bis} shows that the
functions $e(\lambda_{\alpha_k},\lambda_{\alpha_j})$ appearing in the
denominator have been introduced to conveniently rewrite the products
of function $e(\lambda_{\alpha_j},\lambda_{\beta_j})$ in the
numerator. In other words, the pole induced by the functions
$e(\zeta_k,\zeta_j)$ in the denominator of \eqref{topderivation2} are
exactly cancelled by corresponding zeroes in the numerator, hidden in
the products of function $e(\zeta_j,\lambda_{\beta_j})$.

In representation \eqref{topderivation2} for $\psitop{r_1,\dots,r_s}$,
the homogeneous limit in spectral parameters
$\lambda_1,\dots,\lambda_s$, can be done in a straightforward way,
since no new singularity arises as two $\lambda_j$'s assume the same
value. We just set $\lambda_1=\lambda_2=\dots=\lambda_s=\lambda$, and
get
\begin{multline}\label{topderivation3}
\psitop{r_1,\dots,r_s}=
\prod_{k=1}^s \left[a(\lambda,\nu_k)\right]^N
\oint_{C_\lambda}
\cdots \oint_{C_\lambda}
\,\prod_{j=1}^s \,\frac{\left[e(\zeta_j,\lambda)\right]^{r_j-1}
}{
\left[d(\zeta_j,\lambda)\right]^{r_j}}
\prod_{1\leq j<k\leq s}\frac{d(\zeta_k,\zeta_j)}{e(\zeta_k,\zeta_j)}
\\
\times
\frac{ Z_{s}\left(\zeta_1,\dots,\zeta_s;\nu_1,\dots,\nu_s \right)}
{\prod_{j=1}^s\prod_{k=1}^s a(\zeta_j,\nu_k)}
\,\frac{\rmd^s \zeta}{(2\pi \rmi)^s},
\end{multline}
where $C_\lambda$ is a small simple closed counterclockwise contour
around point $\lambda$.  This representation can be related to the
so-called `coordinate wavefunction' representation \eqref{oldrep}
by means of a suitable deformation of the integration contours; we
refer to last part of appendix \ref{app.wave} for details.

Next, we  perform the homogeneous limit in spectral parameters
$\nu_1,\dots,\nu_s$. This is straighforward as well, since all
expressions appearing in \eqref{topderivation3} are regular separately
in this limit. We get
\begin{multline}\label{topderivation4}
\psitop{r_1,\dots,r_s}=
a^{Ns}\oint_{C_\lambda} \cdots \oint_{C_\lambda}
\,\prod_{j=1}^s \,\frac{\left[e(\zeta_j,\lambda)\right]^{r_j-1}
}{
\left[d(\zeta_j,\lambda)\right]^{r_j}}
\prod_{1\leq j<k\leq s}\frac{d(\zeta_k,\zeta_j)}{e(\zeta_k,\zeta_j)}
\\
\times
\frac{Z_{s}\left(\zeta_1,\dots,\zeta_s;0,\dots,0 \right)}{\prod_{j=1}^s
\left[a(\zeta_j)\right]^s}
\frac{\rmd^s \zeta}{(2\pi \rmi)^s}.
\end{multline}

To proceed further we need to use the following identity expressing
the partially inhomogeneous partition function
$Z_N(\lambda_1,\dots,\lambda_N) \equiv
Z_N(\lambda_1,\dots,\lambda_N;0,\dots,0)$ in terms of the generating
function for the one-point boundary correlation function (see
\cite{CP-07b} for further details and proof):
\begin{multline}\label{ZN=hNs}
Z_N(\lambda_1,\dots,\lambda_N)=
Z_N(\lambda,\dots,\lambda)
\prod_{j=1}^N\left(\frac{a(\lambda_j,0)}{a(\lambda,0)}\right)^{N-1}
\\ \times
h_{N,N}(\gamma(\lambda_1-\lambda),\dots,\gamma(\lambda_N-\lambda)).
\end{multline}
Here, the function $\gamma(\xi)$ also depends on $\lambda$ (and $\eta$)
as a parameter and reads:
\begin{equation}\label{change}
\gamma(\xi)\equiv\gamma(\xi;\lambda)=\frac{a(\lambda,0)}{b(\lambda,0)}
\frac{b(\lambda+\xi,0)}{a(\lambda+\xi,0)}.
\end{equation}
Using now the relation \eqref{ZN=hNs} and changing the integration
variables $\zeta_j\mapsto w_j=\gamma(\zeta_j-\lambda)$, we get
\begin{multline}\label{MIRZ1alt}
\psitop{r_1,\dots,r_s}=
Z_s a^{s(N-s)}
\prod_{j=1}^s t^{j-r_j}
\oint_{C_1} \cdots \oint_{C_1}
\prod_{j=1}^s \frac{\left(t^2w_j-2\Delta t +1\right)^{r_j-1}}{(w_j-1)^{r_j}}
\\
\times
\prod_{1\leq j<k\leq s}\frac{w_k-w_j}{t^2 w_j w_k -2 \Delta t w_j+1}
\,h_{s,s}(w_1,\dots,w_s)
\,\frac{\rmd^s w}{(2\pi \rmi)^s}.
\end{multline}
Formula \eqref{MIRZ1alt} is one of our main results here.

Note that our last integral representation \eqref{MIRZ1alt} differs
significantly from \eqref{oldrep_mir}, which is based on the
`coordinate wavefunction' representation \cite{IKR-87}. It appears
that these two representations can be related by a suitable
deformation of integration contours. Such relation is most easily
seen at the level of \eqref{topderivation3}, that is for the
inhomegeneous version of the model, see appendix \ref{app.wave} for
details.

It is worth mentioning that the expression \eqref{MIRZ1alt} also implies,
through crossing symmetry, an analogous $(N-s)$-fold integral
representation for $\psibot{r_1,\dots,r_s}$, depending on the row
configuration through the position $\rbar_1,\dots,\rbar_{N-s}$ of the
$N-s$ down arrows. This representation is given in appendix
\ref{app.dual}, see formula \eqref{MIRZ2alt}. 

Finally, we emphasize that that the two procedures leading from the
inhomogeneous representations \eqref{derivation2} and
\eqref{topderivation1bis} to the multiple integral representations
\eqref{MIRZ2} and \eqref{MIRZ1alt} for $\psibot{r_1,\dots,r_s}$ and
$\psitop{r_1,\dots,r_s}$ are quite different and cannot be
interchanged.

To recapitulate, we have thus in total three different representations
for $\psitop{r_1,\dots,r_s}$ and three for
$\psibot{r_1,\dots,r_s}$. Concerning $\psitop{r_1,\dots,r_s}$, we have
the two representations \eqref{MIRZ1alt} and \eqref{MIRZ1}, besides
the well-known `coordinate wavefunction' representation, see
\eqref{oldrep_mir} for its multiple integral form. Corresponding
representations for $\psibot{r_1,\dots,r_s}$, related to
\eqref{MIRZ1alt} and \eqref{MIRZ1} by crossing symmetry, are given by
\eqref{MIRZ2alt} and \eqref{MIRZ2}, respectively. Each of these
representations appears to depend on the row configuration either
through the $r$'s or through the $\rbar$'s.

\section{Emptiness formation probability}

In this section, we show how the integral representation for the
emptiness formation probability derived in \cite{CP-07b} can be
recovered from the integral representations for
$\psibot{r_1,\dots,r_s}$ and $\psitop{r_1,\dots,r_s}$. The alternative
derivation presented here relies on certain relation involving
antisymmetrization with respect to two set of variables, recently
proved in \cite{CCP-19}.

\subsection{Antisymmetrization relations}

Given a multivariate function, we introduce the antisymmetrizer
\begin{equation}
\Asym_{z_1,\ldots,z_s}f(z_1,\dots,z_s)=
\sum_{\sigma}(-1)^{[\sigma]}
f(z_{\sigma_1},\dots,z_{\sigma_s}), 
\end{equation}
where the sum is taken over the permutations $\sigma:1,\dots, s\mapsto
\sigma_1,\dots,\sigma_s$, with $[\sigma]$ denoting the parity of
$\sigma$.

We discuss here two antisymmetrization relations playing a relevant
role in the calculation of integral representations for the emptiness
formation probability.

The first antisymmetrization relation originates from the following
relation, established and proven by Kitanine \textit{et al.}, see
\cites{KMST-02}, Prop.~C1:
\begin{multline}\label{KMST}
\Asym_{\lambda_1,\ldots,\lambda_s} 
\left[\frac{\prod_{j=1}^{s}\prod_{k=1}^{j-1}a(\lambda_j,\nu_k)
\prod_{k=j+1}^{s}b(\lambda_j,\nu_k)}
{\prod_{1\leq j<k\leq s}^{}e(\lambda_k,\lambda_j)}
\right]
\\
=\frac{\prod_{1\leq j<k\leq s}
  d(\lambda_k,\lambda_j)}{\prod_{j,k=1}^{s}e(\lambda_k,\lambda_j)}
Z_s(\lambda_1,\ldots,\lambda_s;\nu_1,\ldots,\nu_s).
\end{multline}
Here, the functions involved are defined in \eqref{abc},
\eqref{dfunc}, and \eqref{efunc}, and $Z_s$ denotes the
Izergin-Korepin partition function \eqref{ZN} for an $s\times s$
lattice.

We are interested in the particular  case of \eqref{KMST} where
$\nu_j=0$, $j=1,\ldots,s$. We set 
\begin{equation}\label{zjs}
z_j=\gamma(-\lambda_j+\eta), \qquad j=1,\ldots,s,
\end{equation}
where the function $\gamma(\xi)$ is defined in \eqref{change}. We
intend to use \eqref{ZN=hNs}, so it is also convenient to introduce
the notation:
\begin{equation}
u_j=\gamma(\lambda_j-\lambda), \qquad j=1,\ldots,s.
\end{equation}
One has 
\begin{equation}\label{uofz}
u_j=-\frac{z_j-1}{(t^2-2\Delta t)z_j+1}, \qquad j=1,\dots,s,
\end{equation}
where, as above, $\Delta=\cos 2\eta$ and $t\equiv
b(\lambda,0)/a(\lambda,0)$.  Resorting now to \eqref{ZN=hNs}, relation
\eqref{KMST} at $\nu_j=0$, $j=1,\ldots,s$ can be rewritten as the
following antisymmetrization relation:
\begin{multline}\label{identity1}
\Asym_{z_1,\dots,z_s}\left[\prod_{j=1}^{s}\frac{1}{u_j^{s-j}}
\prod_{1\leq j < k \leq s}
(t^2 z_j z_k -2\Delta t z_k+1)
\right]
\\
=(-1)^{\frac{s(s-1)}{2}}\frac{Z_s}{a^{s(s-1)}c^s}
\prod_{1\leq j < k \leq s}(z_k-z_j)
\prod_{j=1}^{s}\frac{1}{u_j^{s-1}}\, h_{s,s}(u_1,\dots,u_s).
\end{multline}
Here, and everywhere below, we assume that $u_j\equiv u(z_j)$, with
the function $u(z_j)$ defined by the right-hand side of
\eqref{uofz}. We refer for more details to \cites{CP-07b}.

The second antisymmetrization relation we wish to discuss reads
\cite{CCP-19}:
\begin{multline}\label{cantini}
\Asym_{x_1,\dots,x_s}\Asym_{y_1,\dots,y_s}\left[
\prod_{j=1}^s  \frac{(x_jy_j)^{s-j}}{1-\prod_{l=1}^j x_ly_l}
\prod_{1\leq j<k\leq s}(x_jx_k-2\Delta x_k+1)(y_jy_k-2\Delta y_k+1)
\right]
\\ =\prod_{j,k=1}^{s}(x_j+y_k-2\Delta x_jy_k)
\det_{1\leq j,k \leq s} \left[\psi(x_j,y_k)\right],
\end{multline}
where 
\begin{equation}
\psi(x,y)=\frac{1}{(1-x y )(x+y-2\Delta x y)}.
\end{equation}
The relation \eqref{cantini} can be proven by induction in $s$, using
the symmetries in the involved variables and comparing singularities
of both sides, along the lines of the proof of the relation
\eqref{KMST} given in \cite{KMST-02}, see appendix C therein.

It is to be mentioned that similar relations appear in connection with
the theory of symmetric polynomials
\cite{KN-99,W-08,BW-16,BWZj-15,P-20}.  Relation \eqref{cantini} does
not seem to be a particular case of any of them, even if sharing the
property that its right-hand side is expressible in terms of the
Izergin-Korepin partition function \eqref{ZN}. Instead, it appears to
extend to the trigonometric case some antisymmetrization relation
originally derived in the rational case by Gaudin, see \cite{G-83},
Appendix B.  Also, \eqref{cantini} generalizes some antisymmetrization
relation given in \cite{TW-08}, in the context of the asymmetric
simple exclusion process.

It is convenient to introduce the notation
\begin{multline}\label{defW}
W_s(x_1,\dots,x_s;y_1,\dots,y_s)
=\frac{\prod_{j,k=1}^{s}
(x_j+y_k-2 \Delta x_jy_k)}{\prod_{1\leq j<k \leq s}(x_k-x_j)
\prod_{1\leq j<k \leq s}(y_k-y_j)}
\\ \times
\det_{1\leq j,k \leq s}  [\psi(x_j,y_k)].
\end{multline}
Observe that  $W_s(x_1,\dots,x_s;y_1,\dots,y_s)$ is a rational 
function of the form
\begin{equation}\label{Wpoly}
W_s(x_1,\dots,x_s;y_1,\dots,y_s)=
\frac{P_s(x_1,\ldots,x_s;y_1,\ldots,y_s)}
{\prod_{j,k=1}^{s}(1-x_jy_k)},
\end{equation}
where $P_s(x_1,\ldots,x_s;y_1,\ldots,y_s)$ is a polynomial of degree
$s-1$ in each variable, separately symmetric under permutations of the
variables within each set.

The function \eqref{defW} is closely related to the Izergin-Korepin
partition function \eqref{ZN}. Indeed, let us set
\begin{equation}\label{xyCantini}
x_j=\frac{a(\lambda_j,\xi+\eta)}{b(\lambda_j,\xi+\eta)}, \qquad
y_j=\frac{a(\xi,\nu_j)}{b(\xi,\nu_j)}, \qquad j=1,\dots,s,
\end{equation}
where $\xi$ is an arbitrary parameter, to be fixed later on. Then, we
have
\begin{equation}\label{cantiniIK}
\det_{1\leq j,k \leq s}\left[\psi(x_j,y_k)\right]
=\frac{1}{c^{3s}}
\prod_{j=1}^s [b(\lambda_j,\xi+\eta)b(\xi,\nu_j)]^2
\det_{1\leq j,k \leq s}\left[\varphi(\lambda_j,\nu_k)\right].
\end{equation}
Plugging this into \eqref{defW} yields
\begin{multline}\label{WsIK}
W_s(x_1,\dots,x_s;y_1,\dots,y_s)=
(-1)^s
\frac{\prod_{j=1}^s b(\lambda_j,\xi+\eta) b(\xi,\nu_j)}{c^{2s}
\prod_{j,k=1}^{s} b(\lambda_j,\nu_k)} 
\\ \times
Z_s(\lambda_1,\ldots,\lambda_s;\nu_1,\ldots,\nu_s).
\end{multline}
Note that here the parameter $\xi$ enters only the prefactor and not
the Izergin-Korepin partition function.

Consider now \eqref{WsIK} in the partially homogeneous limit where
$\nu_j\to 0$, $j=1,\ldots,s$. To make contact with our previous
discussion let us also identify $\xi=\lambda$. By comparison of
\eqref{xyCantini} with \eqref{zjs}, in the limit we get
\begin{equation}
x_j=t z_j,\quad y_j=t^{-1},\qquad j=1,\ldots,s.
\end{equation}  
Recalling \eqref{ZN=hNs}, we thus obtain
\begin{equation}\label{Whom}
W_s(t z_1,\dots,t z_s;t^{-1},\dots,t^{-1})
=\frac{(-1)^s Z_s}{c^s b^{s(s-1)}}
\prod_{j=1}^s \frac{1}{(z_j-1)u_j^{s-1}}\,
h_{s,s}(u_1,\dots,u_s),
\end{equation}
where, as usual, $u_j$'s and $z_j$'s are related by \eqref{uofz}.  

Another result of interest concerns the evaluation of the quantity
$P_s(x_1,\dots,x_s;$ $x_1^{-1},\dots,x_s^{-1})$, see \eqref{Wpoly}. From
\eqref{Wpoly}, \eqref{xyCantini}, and \eqref{WsIK}, it follows that:
\begin{multline}\label{Psxy}
  P_s(x_1,\dots,x_s;y_1,\dots,y_s)=
  c^{s^2-2s}\prod_{j=1}^s
  \frac{1}{[b(\lambda_j,\xi+\eta)b(\xi,\nu_j)]^{s-1}}
  \\
  \times
Z_s(\lambda_1,\ldots,\lambda_s;\nu_1,\ldots,\nu_s).
\end{multline}
Setting now $\nu_j=\lambda_j-\eta$, $j=1,\dots,s$, we get
\begin{multline}\label{Psxx}
  P_s(x_1,\dots,x_s;x_1^{-1},\dots,x_s^{-1})=
  c^{s^2-2s}\prod_{j=1}^s
  \frac{1}{[b(\lambda_j,\xi+\eta)b(\xi,\lambda_j-\eta)]^{s-1}}
  \\
  \times
Z_s(\lambda_1,\ldots,\lambda_s;\lambda_1-\eta,\ldots,\lambda_s-\eta).
\end{multline}
The last line is easily evaluated thanks to the recursion relation for
the inhomogeneous partition function \cite{K-82}. We have
\begin{equation}
  Z_s(\lambda_1,\ldots,\lambda_s;\lambda_1-\eta,\ldots,\lambda_s-\eta)=
  c^s\prod_{\substack{j,k=1\\j\ne k}}^s\sin(\lambda_j-\lambda_k+2\eta).
\end{equation}
Reexpressing now the right-hand side of \eqref{Psxx} in terms of
variables $x_1,\dots,x_j$, we finally obtain
\begin{equation}\label{psxx2}
P_s(x_1,\dots,x_s;x_1^{-1},\dots,x_s^{-1})=\prod_{j=1}^s\frac{1}{x_j^{s-1}}
\prod_{\substack{j,k=1\\j\ne k}}^s(x_jx_k-2\Delta x_j+1),
\end{equation}
which will turn out useful below.

\subsection{Known integral representations}
  
In \cite{CP-07b} various representations has been worked out for the
emptiness formation probability $F_N^{(r,s)}$. In particular, using
the Yang-Baxter commutation relations, and next performing the
homogeneous limit, the following representation in terms of the
orthogonal polynomials was obtained:
\begin{multline}\label{orthEFP}
F_N^{(r,s)}=(-1)^s
\begin{vmatrix}
K_{N-s}(\partial_{\eps_1}) &\hdots & K_{N-s}(\partial_{\eps_s})\\
\hdotsfor{3} \\
K_{N-1}(\partial_{\eps_1}) &\hdots & K_{N-1}(\partial_{\eps_s})
\end{vmatrix}
\prod_{j=1}^{s}\left\{
\frac{[\omega(\eps_j)]^{N-r}}{[\omega(\eps_j)-1]^N} \right\}
\\ \times
\prod_{1\leq j<k \leq s}^{} \frac{ [1-\tilde\omega(\eps_j)][\omega(\eps_k)-1]}
{[\tilde\omega(\eps_j)\omega(\eps_k)-1]}
\Bigg|_{\eps_1,\ldots,\eps_s=0}.
\end{multline} 
Here, we use the same notations as in section \ref{Sect41}. In
particular, $K_n(x)$ denote the orthogonal polynomials \eqref{Knx},
associated to the Hankel matrix \eqref{varphi}, and the functions
$\omega(\eps)$ and $\tilde\omega(\eps)$ are defined in \eqref{omegas}.

The identity \eqref{claim} when applied to \eqref{orthEFP}, yields for
the emptiness formation probability the following multiple integral
representation:
\begin{multline}\label{efpMIR1}
F_N^{(r,s)} = (-1)^s
\oint_{C_0}^{} \cdots \oint_{C_0}^{}
\prod_{j=1}^{s}\frac{[(t^2-2\Delta t)z_j+1]^{s-j}}{z_j^r(z_j-1)^{s-j+1}}\,
\\ \times
\prod_{1\leq j<k \leq s}^{} \frac{z_j-z_k}{t^2z_jz_k-2\Delta t z_j+1}\;
h_{N,s}(z_1,\dots,z_s)
\,\frac{\rmd^s z}{(2\pi \rmi)^s}.
\end{multline}
Here, $C_0$ denotes, as before, a small simple anticlockwise oriented
contour around the point $z=0$, and the function
$h_{N,s}(z_1,\dots,z_s)$ is defined in \eqref{hNs}.

Note that the integrand in  \eqref{efpMIR1} is not symmetric with
respect to the permutation of the integration variables. However, the
antisymmetrization relation \eqref{identity1} allows to write down
the essentially equivalent representation
\begin{multline}\label{efpMIR2}
F_N^{(r,s)}=\frac{(-1)^sZ_s}{s!\,a^{s(s-1)}c^s}
\oint_{C_0}^{} \cdots \oint_{C_0}^{}
\prod_{j=1}^{s} \frac{[(t^2-2\Delta t)z_j+1]^{s-1}}{z_j^r(z_j-1)^s}\,
\\ \times
\prod_{\substack{j,k=1\\ j\ne k}}^{s} \frac{z_k-z_j}{t^2 z_jz_k-2\Delta t z_j +1}
h_{N,s}(z_1,\dots,z_s)h_{s,s}(u_1,\dots,u_s)
\,\frac{\rmd^s z}{(2\pi \rmi)^s}.
\end{multline}
Here, $u_j$'s are given in terms of $z_j$'s by \eqref{uofz}.  The
representation \eqref{efpMIR2} with the symmetric integrand has been proved
of importance, for example, in the evaluation of the phase separation
curves of the model \cite{CP-08,CP-09}.

Given integral representation \eqref{efpMIR2} for the emptiness
formation probability, a natural question concerns the possibility of
deriving it by suitably combining the integral expressions obtained
above for $\psitop{r_1,\dots,r_s}$ and $\psibot{r_1,\dots,r_s}$.  As
we will show below, the answer is affermative.

\subsection{An alternative and simpler derivation}\label{sec.recovering}

We propose here an alternative derivation of \eqref{efpMIR2}, with
respect to the one originally proposed in \cite{CP-07b}. Here we start
from the relation \eqref{efp}, and substitute the integral
representations \eqref{oldrep_mir} and \eqref{MIRZ2} for
$\psitop{r_1,\dots,r_s}$ and $\psibot{r_1,\dots,r_s}$ in the
expression for the row configuration probability \eqref{defHNs}.  An
essential role in the derivation is played by relation
\eqref{cantini}.

For convenience, we change the integration variables $z_j\mapsto
x_j/t$, $j=1,\dots,s$ in \eqref{MIRZ2}, that yields
\begin{multline}\label{zbotcov}
\psibot{r_1,\dots,r_s}=
\frac{Z_N}{a^{s(N-1)}c^s}
\oint_{C_0}^{} \cdots \oint_{C_0}^{}
\prod_{j=1}^s\frac{1}{x^{r_j}_j}
\prod_{1\leq j <k \leq s}\frac{x_k-x_j}{x_j x_k -2\Delta  x_j+1}
\\ \times
h_{N,s}\left(\frac{x_1}{t},\dots,\frac{x_s}{t}\right)
\frac{\rmd^s x}{(2\pi \rmi)^s},
\end{multline}
and also we change $w_j\mapsto 1/(ty_j)$ in \eqref{oldrep_mir}, 
that yields
\begin{multline}\label{ztopoldcov}
\psitop{r_1,\dots,r_s}
=c^s a^{s(N-1)}
\oint_{C_{1/t}}^{} \cdots \oint_{C_{1/t}}^{} \prod_{j=1}^s
\frac{1}{(ty_j-1)^s y_j^{r_j+s-1}}
\\ \times
\prod_{1\leq j<k\leq s} \left[(y_k-y_j) (y_j y_k -2\Delta y_k+1)\right]
\frac{\rmd^s y}{(2\pi\rmi)^s}.
\end{multline}
Then, inserting into  \eqref{defHNs} and \eqref{efp}, we have
\begin{multline}\label{efpdoubleMIR}
F_N^{(r,s)}=
\oint_{C_{1/t}}^{} \cdots \oint_{C_{1/t}}^{}
\frac{\rmd^s y}{(2\pi \rmi)^s}
\oint_{C_0}^{} \cdots \oint_{C_0}^{}
\prod_{j=1}^s \frac{1}{y_j^{s-1}(ty_j-1)^s}
\\\times
\prod_{1\leq j<k\leq s}
\frac{(y_k-y_j)(y_j y_k -2\Delta y_k +1)(x_k-x_j)}{x_j x_k -2 \Delta x_j +1}
h_{N,s}\left(\frac{x_1}{t},\dots,\frac{x_s}{t}\right)
\\\times
\sum_{1\leq r_1 < r_2 < \dots < r_s \leq r}
\prod_{j=1}^s \frac{1}{(x_jy_j)^{r_j}}
\frac{\rmd^s x}{(2\pi \rmi)^s}.
\end{multline}
To prove that this representation indeed reduces  to \eqref{efpMIR2}, 
one has to perform  the multiple sum and evaluate $s$ integrations.  

First, let us focus on the multiple sum in \eqref{efpdoubleMIR}. It is
clear that, due to the integrations around the points $x_j=0$,
$j=1,\ldots,s$, one can extend the sum over the values $1\leq
r_1<r_2\cdots<r_s\leq r$ to the values $-\infty<r_1<r_2\cdots<r _s\leq
r$, obtaining:
\begin{equation}\label{summation}
\sum_{-\infty< r_1 < r_2 < \dots < r_s \leq r}
\prod_{j=1}^s \frac{1}{(x_jy_j)^{r_j}}
=\prod_{j=1}^s \frac{1}{(x_jy_j)^{r-s+j}\big(1-\prod_{l=1}^j x_ly_l\big)}.
\end{equation}
Hence, \eqref{efpdoubleMIR} simplifies to
\begin{multline}\label{efpdoubleMIR2}
F_N^{(r,s)}=
\oint_{C_{1/t}}^{} \cdots \oint_{C_{1/t}}^{}
\frac{\rmd^s y}{(2\pi \rmi)^s}
\oint_{C_0}^{} \cdots \oint_{C_0}^{}
\prod_{j=1}^s \frac{1}{(ty_j-1)^s y_j^{r+j-1}x_j^{r-s+j}}
\\ \times
\prod_{j=1}^{s}\frac{1}{(1-\prod_{l=1}^{j} x_l y_l)}
\prod_{1\leq j<k\leq s}{}
\frac{(y_k-y_j)(y_j y_k -2\Delta  y_k +1)(x_k-x_j)}{x_j x_k -2 \Delta x_j +1}
\\ \times 
h_{N,s}\left(\frac{x_1}{t},\dots,\frac{x_s}{t}\right)
\frac{\rmd^s x}{(2\pi \rmi)^s}.
\end{multline}

To perform the integrations, we first observe that the form of the
integrand in \eqref{efpdoubleMIR2} allows us to apply the
antisymmetrization relation \eqref{cantini}. We obtain
\begin{multline}\label{efpdoubleMIR3}
F_N^{(r,s)}=\frac{1}{(s!)^2}
\oint_{C_{1/t}}^{} \cdots \oint_{C_{1/t}}^{}
\frac{\rmd^s y}{(2\pi \rmi)^s}
\oint_{C_0}^{} \cdots \oint_{C_0}^{}
\prod_{j=1}^s \frac{1}{x_j^r(ty_j-1)^s y_j^{r+s-1}}
\\\times 
\prod_{1\leq j<k\leq s}
\frac{(x_k-x_j)^2(y_k-y_j)^2}{(x_j x_k -2 \Delta  x_j +1)
(x_j x_k -2 \Delta  x_k +1)}
\\ \times
W_s(x_1,\dots,x_s;y_1,\dots,y_s)
h_{N,s}\left(\frac{x_1}{t},\dots,\frac{x_s}{t}\right)
\frac{\rmd^s x}{(2\pi \rmi)^s},
\end{multline}
where we have used the notation \eqref{defW}.

To evaluate the integrals over the variables $y_1,\dots,y_s$, we  resort
to  the following identity:
\begin{equation}\label{symmint}
\oint_{C_{w_1,\dots,w_s}}\dots\oint_{C_{w_1,\dots,w_s}}
\frac{\prod_{\substack{j,k=1\\j\ne k}}^s(y_j-y_k)}{\prod_{1\leq j,k\leq s }(y_j-w_k)}
\Phi(y_1,\dots,y_s)\frac{\rmd^s y} {(2\pi\rmi)^s} = s!
\Phi(w_1,\dots,w_s),
\end{equation}
where $C_{w_1,\dots,w_s}$ is a simple closed counterclockwise contour
in the complex plane, enclosing points $w_1,\dots,w_s$, and
$\Phi(y_1,\dots,y_s)$ is a generic symmetric function, analytic in
each of its variables within a domain containing $C_{w_1,\dots,w_s}$.
In our case $w_j=t^{-1}$, $j=1,\dots,s$, and we obtain
\begin{multline}\label{efpMIR2recovered}
F_N^{(r,s)}=\frac{t^{s(r-1)}}{s!}
\oint_{C_0}^{} \cdots \oint_{C_0}^{}
\prod_{j=1}^s \frac{1}{x_j^r}
\prod_{\substack{j,k=1\\ j\ne k}}
\frac{x_j-x_k}{x_j x_k -2 \Delta x_j +1}
\\ \times
W_s(x_1,\dots,x_s;t^{-1},\dots,t^{-1})
h_{N,s}\left(\frac{x_1}{t},\dots,\frac{x_s}{t}\right)
\frac{\rmd^s x}{(2\pi \rmi)^s}.
\end{multline}
Finally, changing back the integration variables $x_j\mapsto t z_j$,
$j=1,\dots,s$, and using \eqref{Whom}, we recover \eqref{efpMIR2}.

\section{Another  representation for the emptiness formation probability}

In this section we derive an alternative representation for the
emptiness formation probability, which supplements the already known
representation \eqref{efpMIR2}. Crucial ingredients in such derivation
are: \textit{i)} the use of relation \eqref{efpn} rather than
\eqref{efp} to express the emptiness formation probability in terms of
$\psitop{r_1,\dots,r_s}$ and $\psibot{r_1,\dots,r_s}$, and
\textit{ii)} the use of representation \eqref{MIRZ1alt}, rather than
\eqref{oldrep_mir}, for $\psitop{r_1,\dots,r_s}$. Although the
starting point is quite different, the derivation has some
similarities with that of section \ref{sec.recovering}, in particular,
we will again use the antisymmetrization relation \eqref{cantini}.

\subsection{Performing summations}

Our starting point is relation \eqref{efpn}, with
$H_N^{(1,\dots,s,r_{s+1},\dots,r_{s+n})}$ given by \eqref{defHNs}. As
for $\psitop{1,\dots,s,r_{s+1},\dots,r_{s+n}}$, we resort to
representation \eqref{MIRZ1alt}, which, in the present setup, reads
\begin{multline}\label{ztopnewcov2}
\psitop{r_1,\dots,r_{s+n}}=
Z_{s+n} a^{(s+n)(N-s-n)}
\\ \times
\oint_{C_0} \cdots \oint_{C_0}
\prod_{j=1}^{s+n} \frac{1}{w_j^{r_j}(1-tw_j)}
\prod_{1\leq j<k\leq s+n}\frac{w_k-w_j}{w_j w_k -2 \Delta  w_j+1}
\\ \times
h_{s+n,s+n}\left(\frac{(2\Delta t -1)w_1-t}{t(tw_1 -1)},\dots,
\frac{(2\Delta t -1)w_{s+n}-t}{t(tw_{s+n} -1)}\right)
\frac{\rmd^{s+n} w}{(2\pi \rmi)^{s+n}}.
\end{multline}
Having $r_j=j$, $j=1,\dots,s$, integration over variables
$w_1,\dots,w_s$, in this order, is easily done, since each time the
pole at origin is of order one.  We thus have
\begin{multline}\label{ztopnewcov3}
\psitop{1,\dots,s,s+l_1,\dots,s+l_n}=
Z_{s+n} a^{(s+n)(N-s-n)}
\\ \times
\oint_{C_0} \cdots \oint_{C_0}
\prod_{j=1}^{n} \frac{1}{w_j^{l_j}(1-tw_j)}
\prod_{1\leq j<k\leq n}\frac{w_k-w_j}{w_j w_k -2 \Delta  w_j+1}
\\ \times
h_{s+n,n}\left(\frac{(2\Delta t -1)w_1-t}{t(tw_1 -1)},\dots,
\frac{(2\Delta t -1)w_{n}-t}{t(tw_{n} -1)}\right)
\frac{\rmd^{n} w}{(2\pi \rmi)^{n}},
\end{multline}
where we have used relation \eqref{hns1}, and we have relabelled
$r_{s+j}\to l_j:=r_{s+j}-s$, $j=1,\dots,n$.  Note that we would have
not been able to evaluate these first $s$ integration if we had
started from expression \eqref{oldrep_mir}, rather than
\eqref{MIRZ1alt}, for $\psitop{1,\dots,s,s+l_1,\dots,s+l_n}$.

We turn now to $\psibot{1,\dots,s,r_{s+1},\dots,r_{s+n}}$, and resort
to representation \eqref{MIRZ2}, or equivalently, up to  change of
the integration variables, to \eqref{zbotcov}, which, in the present
setup, reads:
\begin{multline}\label{zbotcov2}
\psibot{r_1,\dots,r_{s+n}}=
\frac{Z_N}{a^{(s+n)(N-1)}c^{s+n}}
\oint_{C_0}^{} \cdots \oint_{C_0}^{}
\prod_{j=1}^{s+n}\frac{1}{z^{r_j}_j}
\\ \times
\prod_{1\leq j <k \leq {s+n}}\frac{z_k-z_j}{z_j z_k -2\Delta  z_j+1}
h_{N,s+n}\left(\frac{z_1}{t},\dots,\frac{z_{s+n}}{t}\right)
\frac{\rmd^{s+n} z}{(2\pi \rmi)^{s+n}}.
\end{multline}
Again, having $r_j=j$, $j=1,\dots,s$, the first $s$ integrations are
easily done, with the result
\begin{multline}\label{zbotcov3}
\psibot{1,\dots,s,s+l_1,\dots,s+l_{n}}=
\frac{Z_{N-s}a^{s(N-s)}}{c^{n}a^{n(N-1)}} \oint_{C_0}^{} \cdots
\oint_{C_0}^{} \prod_{j=1}^{n}\frac{1}{z^{l_j}_j}
\\ \times
\prod_{1\leq j <k \leq {n}}\frac{z_k-z_j}{z_j z_k -2\Delta  z_j+1}
h_{N-s,n}\left(\frac{z_1}{t},\dots,\frac{z_{n}}{t}\right)
\frac{\rmd^{n} z}{(2\pi \rmi)^{n}},
\end{multline}
where we have used the relation \eqref{hns0}.

Substituting now \eqref{ztopnewcov3} and \eqref{zbotcov3} into
\eqref{efpn}, see \eqref{defHNs}, we have
\begin{multline}
F_N^{(s+n,s)}= \frac{Z_{s+n}Z_{N-s}a^{2s(N-s-n)}}{Z_Nc^na^{n(n-1)}}
\oint_{C_0}\cdots\oint_{C_0} \prod_{j=1}^n \frac{1}{1-tw_j}
\\ \times
\sum_{1\leq l_1<\dots<l_n\leq N-s}\prod_{j=1}^n\frac{1}{(w_jz_j)^{l_j}}
\prod_{1\leq j<k\leq n}
\frac{(w_k-w_j)(z_k-z_j)}{(w_jw_k-2\Delta w_j+1)(z_jz_k-2\Delta  z_j+1)}
\\ \times
h_{s+n,n}\left(\frac{(2\Delta t -1)w_1-t}{t(tw_1 -1)},\dots,
\frac{(2\Delta t -1)w_{n}-t}{t(tw_{n} -1)}\right)
\\ \times
h_{N-s,n}\left(\frac{z_1}{t},\dots,\frac{z_{n}}{t}\right)
\frac{\rmd^nw}{(2\pi\rmi)^n}\frac{\rmd^nz}{(2\pi\rmi)^n}.
\end{multline}
Observing that, similarly to  section \ref{sec.recovering}, the integral
over $z_j$ vanishes for $l_j<0$, we can extend the sum over all
negative values. Using then formula \eqref{summation}, we obtain
\begin{multline}\label{nefp1}
F_N^{(s+n,s)}= \frac{Z_{s+n}Z_{N-s}a^{2s(N-s-n)}}{Z_Nc^na^{n(n-1)}}
\\ \times
\oint_{C_0}\cdots\oint_{C_0} \prod_{j=1}^n
 \frac{1}{(1-tw_j)(w_jz_j)^{N-s-n+j}\left(1-\prod_{l=1}^jw_lz_l\right)}
\\ \times
\prod_{1\leq j<k\leq n}
\frac{(w_k-w_j)(z_k-z_j)}{(w_jw_k-2\Delta w_j+1)(z_jz_k-2\Delta  z_j+1)}
\\ \times
h_{s+n,n}\left(\frac{(2\Delta t -1)w_1-t}{t(tw_1 -1)},\dots,
\frac{(2\Delta t -1)w_{n}-t}{t(tw_{n} -1)}\right)
\\ \times
h_{N-s,n}\left(\frac{z_1}{t},\dots,\frac{z_{n}}{t}\right)
\frac{\rmd^nw}{(2\pi\rmi)^n}\frac{\rmd^nz}{(2\pi\rmi)^n}.
\end{multline}
We have thus performed the summations in \eqref{efpn}.

\subsection{Deforming integration contours}\label{sec.deforming}

We want now to address the question of performing integration with
respect to variables $z_1,\dots,z_n$. First of all, along the lines of
section \ref{sec.recovering}, let us symmetrize the integrand in
\eqref{nefp1} by resorting to relation \eqref{cantini}. We obtain
\begin{multline}\label{nefp2}
F_N^{(s+n,s)}= \frac{Z_{s+n}Z_{N-s}a^{2s(N-s-n)}}{(n!)^2 Z_Nc^na^{n(n-1)}}
\oint_{C_0}\cdots\oint_{C_0} \prod_{j=1}^n \frac{1}{1-tw_j}
\\ \times
\prod_{j=1}^n \frac{1}{(w_jz_j)^{N-s}} 
\prod_{\substack{j,k=1\\j\ne k}}^n
\frac{(w_k-w_j)(z_k-z_j)}{(w_jw_k-2\Delta w_j+1)(z_jz_k-2\Delta  z_j+1)}
\\ \times
h_{s+n,n}\left(\frac{(2\Delta t -1)w_1-t}{t(tw_1 -1)},\dots,
\frac{(2\Delta t -1)w_{n}-t}{t(tw_{n} -1)}\right)
\\ \times
\frac{P_n(w_1,\dots,w_n;z_1,\dots,z_n)}{\prod_{1\leq j,k\leq n}(1-w_jz_k)}
h_{N-s,n}\left(\frac{z_1}{t},\dots,\frac{z_{n}}{t}\right)
\frac{\rmd^nw}{(2\pi\rmi)^n}\frac{\rmd^nz}{(2\pi\rmi)^n},
\end{multline}
where we have used the notation \eqref{Wpoly}.

Turning now to the integration with respect to variables
$z_1,\dots,z_n$, we observe that the corresponding integration
contours $C_0$ can be deformed into new contours
$\Cws:=C_{1/w_1}\cup\dots\cup C_{1/w_n}$, enclosing the poles at
$1/w_j$, $j=1,\dots,n$, induced by the factor
${\prod_{1\leq j,k\leq n}(1-w_jz_k)}$ in the
denominator of \eqref{nefp2}, without changing the result.

The crucial ingredient permitting such deformation of contours is that
the poles induced by the double product in the second line of
\eqref{nefp2} give a vanishing contribution to the integral. This
should not come as a surprise, since such double product is the
remnant of analogous double products appearing in \eqref{derivation2}
and \eqref{topderivation2}. As already commented thereafter, the poles
associated to such double products are exactly compensated by
corresponding zeroes.  However, once the homogeneous limit is
performed, one loose track of this fact, and the mechanism of
cancellation becomes quite subtle. A direct and detailed description
is therefore appropriate.

To start with, let us focus on variable $z_n$. The poles induced by
the double product in the second line of \eqref{nefp2} can be divided
into a first set of $n-1$ poles at positions $(2\Delta z_j -1)/z_j$,
$j=1,\dots,n-1$, and a second set of $n-1$ poles at positions
$1/(2\Delta-z_j)$, $j=1,\dots,n-1$. Concerning the first set of poles,
recalling the symmetry of the integrand under interchange of the
$z_j$'s, let us focus for definiteness on the pole corresponding to
$j=1$, and show that the residue of the integrand in \eqref{nefp2} at
$z_n=(2\Delta z_1 -1)/z_1$ vanishes upon integration with respect to
variable $z_1$ over contour $C_0$.

Indeed, let us inspect the small $z_1$ behaviour of the residue of the
integrand at $z_n=(2\Delta z_1 -1)/z_1$. Such behaviour results from
the contribution of four terms. Let us analyze them in turn.
For the first relevant term, it is easily checked that
\begin{equation}
  \prod_{j=1}^n\frac{1}{z_j^{N-s}}
  \Big|_{z_n=\frac{2\Delta z_1    -1}{z_1}} \sim 1,\qquad z_1\to 0.
  \end{equation}
The next term, which contains the considered pole, requires a careful
but neverthless straightforward calculation, that gives
\begin{equation}
  \lim_{z_n\to\frac{2\Delta z_1-1}{z_1}} \left(z_n-\frac{2\Delta
    z_1-1}{z_1}\right) \prod_{\substack{j,k=1\\j\ne k}}^n
  \frac{z_k-z_j}{z_jz_k-2\Delta z_j+1}
  \sim \frac{1}{z_1^2},\qquad z_1\to 0.
\end{equation}
As for the third relevant term, recalling that
$P_n(w_1,\dots,w_n;z_1,\dots,z_n)$ is a polynomial of degree $n-1$ in
each of its variables, it follows that
\begin{equation}\label{behaviour2}
\frac{P_n(w_1,\dots,w_n;z_1,\dots,z_n)}{\prod_{1\leq j,k\leq n}(1-w_jz_k)}
\Big|_{z_n=\frac{2\Delta  z_1 -1}{z_1}}
\sim z_1,\qquad z_1\to 0.
\end{equation}
Finally, concerning the contribution of $h_{N,n}(z_1/t,\dots,z_n/t)$,
its calculation is nontrivial. However we can resort to the following
property
\begin{equation}\label{behaviour1}
 h_{N,n}\left(\frac{z_1}{t},\dots,\frac{z_n}{t}\right)
 \Big|_{z_n=\frac{2\Delta z_1 -1}{z_1}} \sim z_1,\qquad z_1\to 0,
\end{equation}
which follows from the determinantal structure of the Izergin-Korepin
partition function, and has been proven in \cite{CPS-16}, see appendix
therein.

In all, it follows that the residue of the integrand in \eqref{nefp2}
at $z_n=(2\Delta z_1 -1)/z_1$ is $O(1)$ as $z_1\to 0$, and thus
vanishes upon integration with respect to variable $z_1$ over contour
$C_0$.  Due to the simmetry of the integrand under interchange of the
$z_j$'s, the same holds for variables $z_2,\dots,z_{n-1}$. The
integration contour for $z_n$ can thus be deformed from $C_0$ to a new
contour $C'$ enclosing the origin and the poles at $z_n=(2 \Delta
z_j-1)/z_j$, $j=1,\dots,n-1$.

Inspecting now the large $|z_n|$ behaviour of the integrand, we
observe that the double product in the second line is $O(1)$. We also
have
\begin{equation}
  P_n(w_1,\dots,w_n;z_1,\dots,z_n) \sim |z_n|^{n-1}, \qquad |z_n|\to\infty,
\end{equation}
and
\begin{equation}
  h_{N-s,n}\left(\frac{z_1}{t},\dots,\frac{z_n}{t}\right) \sim |z_n|^{N-s-1},
  \qquad  |z_n|\to\infty.
\end{equation}
As a result, the whole integrand is $O(1/z_n^2)$ as $|z_n|\to \infty$,
and the integration over $z_n$ along a large contour vanishes.  It
follows that we can deform the integration contour over variable $z_n$
from $C'$, defined in the previous paragraph to a new contour $C''$
enclosing the poles at $z_n=1/(2\Delta-z_j)$, $j=1,\dots,n-1$, and the
poles at $z_n=1/w_l$, $l=1,\dots, n$, induced by the factor
${\prod_{1\leq j,k\leq n}(1-w_jz_k)}$ in the denominator of
\eqref{nefp2}.

Implementing the same procedure for other integration variables as
well, we conclude that for each $z_j$, the corresponding integration
contour can be deformed into a contour $C''$ enclosing only the poles
at $z_j=1/w_l$, $l=1,\dots, n$, induced by the denominator in the last
line of \eqref{nefp2}, and the poles at $z_j=1/(2 \Delta - z_l)$,
$l=1,\dots, j-1$.

Now, concerning the poles at $z_n=1/(2 \Delta - z_j)$, $j=1,\dots, n$,
it appears that their contributions vanishes as well. Indeed,
focussing again, for definiteness, on the pole corresponding to $j=1$,
$z_n=1/(2\Delta-z_1)$, consider the following crucial property of the
multivariate polynomial $P_n(w_1,\dots,w_n;z_1,\dots,z_n)$,
\begin{equation}\label{behaviour3}
  P_n(w_1,\dots,w_n;z_1,\dots,z_n)\Big|_{z_n=\frac{1}{2\Delta - z_1}}\sim
  (1-w_kz_1), \qquad z_1\to\frac{1}{w_k},\qquad 
    k=1,\dots,n,
  \end{equation}
whose proof goes along the lines of \eqref{behaviour1}. This property
implies that, after evaluation of the residue of the integrand in
\eqref{nefp2} at $z_n=1/(2 \Delta - z_1)$, the poles at $z_1=1/w_k$,
$k=1,\dots,n$, are all cancelled. It other words, the residue of the
integrand at $z_n=1/(2 \Delta - z_j)$, when subsequently integrated
with respect to $z_j$ over contour $\Cws$, gives a vanishing
contribution.  It follows that we can deform the integration contour
over variable $z_n$ from $C''$ to shrink it down to $\Cws$.

Implementing the same procedure for other integration variables as
well, we have eventually shown that we can deform each of the
integration contours for variables $z_1,\dots,z_n$ from $C_0$ to a new
contour $\Cws$ enclosing the poles at $1/w_k$, $k=1,\dots,n$.  In
other words, \eqref{nefp2} can be rewritten as
\begin{multline}\label{nefp2bis}
F_N^{(s+n,s)}= \frac{Z_{s+n}Z_{N-s}a^{2s(N-s-n)}}{(n!)^2 Z_Nc^na^{n(n-1)}}
\\ \times
\oint_{C_0}\cdots\oint_{C_0} \frac{\rmd^nw}{(2\pi\rmi)^n}
\oint_{\Cws}\cdots\oint_{\Cws}
\prod_{j=1}^n \frac{1}{(1-tw_j)(w_jz_j)^{N-s}} 
\\ \times
\prod_{\substack{j,k=1\\j\ne k}}^n
\frac{(w_k-w_j)(z_k-z_j)}{(w_jw_k-2\Delta w_j+1)(z_jz_k-2\Delta  z_j+1)}
\\ \times
h_{s+n,n}\left(\frac{(2\Delta t -1)w_1-t}{t(tw_1 -1)},\dots,
\frac{(2\Delta t -1)w_{n}-t}{t(tw_{n} -1)}\right)
\\ \times
\frac{P_n(w_1,\dots,w_n;z_1,\dots,z_n)}{\prod_{1\leq j,k\leq n}(1-w_jz_k)}
h_{N-s,n}\left(\frac{z_1}{t},\dots,\frac{z_{n}}{t}\right)
\frac{\rmd^nz}{(2\pi\rmi)^n}.
\end{multline}
The main benefit of this last expression is evident: the integrations
with respect to the variables $z_1,\dots,z_n$ involve now only residues
at simple poles.

\subsection{Performing integrations}

The integration with respect to variables $z_1,\dots,z_n$ in
\eqref{nefp2bis} can be  easily performed. Indeed, resorting to
\eqref{symmint}, we obtain
\begin{multline}\label{nefp3}
F_N^{(s+n,s)}= \frac{Z_{s+n}Z_{N-s}a^{2s(N-s-n)}}{n! Z_Nc^na^{n(n-1)}}
\oint_{C_0}\cdots\oint_{C_0} \prod_{j=1}^n \frac{w_j^{n-2}}{1-tw_j}
\\ \times
\prod_{\substack{j,k=1\\j\ne k}}^n
\frac{(w_k-w_j)}{(w_jw_k-2\Delta w_j+1)(w_jw_k-2\Delta  w_j+1)}
\\ \times
h_{s+n,n}\left(\frac{(2\Delta t -1)w_1-t}{t(tw_1 -1)},\dots,
\frac{(2\Delta t -1)w_{n}-t}{t(tw_{n} -1)}\right)
\\ \times
P_n(w_1,\dots,w_n;w_1^{-1},\dots,w_n^{-1})
h_{N-s,n}\left(\frac{1}{tw_1},\dots,\frac{1}{tw_n}\right)
\frac{\rmd^nw}{(2\pi\rmi)^n}.
\end{multline}
Recalling \eqref{psxx2}, we get
\begin{multline}
F_N^{(s+n,s)}= \frac{Z_{s+n}Z_{N-s}a^{2s(N-s-n)}}{n! Z_Nc^na^{n(n-1)}}
\oint_{C_0}\cdots\oint_{C_0} \prod_{j=1}^n \frac{1}{w_j(1-tw_j)}
\\ \times
\prod_{\substack{j,k=1\\j\ne k}}^n
\frac{w_k-w_j}{w_jw_k-2\Delta w_j+1}
\,
h_{N-s,n}\left(\frac{1}{w_1t},\dots,\frac{1}{w_nt}\right)
\\ \times
h_{s+n,n}\left(\frac{(2\Delta t -1)w_1-t}{t(tw_1 -1)},\dots,
\frac{(2\Delta t -1)w_{n}-t}{t(tw_{n} -1)}\right)
\frac{\rmd^nw}{(2\pi\rmi)^n}.
\end{multline}
Finally, performing the change of variables $w_j\mapsto 1/(t z_j)$,
$j=1,\dots,n$, we arrive at the following expression
\begin{multline}\label{nefp_final}
F_N^{(s+n,s)}= \frac{Z_{s+n}Z_{N-s}a^{2s(N-s-n)}t^{n(n-1)}}{n! Z_Nc^na^{n(n-1)}}
\oint_{C_1}\cdots\oint_{C_1} \prod_{j=1}^n \frac{1}{z_j-1}
\\ \times
\prod_{\substack{j,k=1\\j\ne k}}^n
\frac{z_j-z_k}{t^2z_jz_k-2\Delta t z_j+1}
\,
h_{N-s,n}\left(z_1,\dots,z_n\right)
\\ \times
h_{s+n,n}\left(\frac{t^2z_1-2\Delta t +1}{t^2(z_1-1)},\dots,
\frac{t^2z_n-2\Delta t +1}{t^2(z_n-1)}\right)
\frac{\rmd^n z}{(2\pi\rmi)^n},
\end{multline}
where we have shrunk each of the $n$ integration contours from a very
large one down to a small contour enclosing only the pole at $z_j=1$,
$j=1,\dots, n$, thus ignoring the contribution of the poles in the
double product. Once again, it can be shown that the total contribution
of such poles indeed vanishes.

We consider the multiple integral representation \eqref{nefp_final} as
one of the main results of the present paper. At variance with the
previously known representation \eqref{efpMIR2}, the number of
integrations is $n=r-s$, that is the lattice distance of the point
$(r,s)$ from the antidiagonal, rather than $s$, the lattice distance
from the top boundary. This could be useful to investigate the
behaviour of the emptiness formation probability in the so-called
Hamiltonian limit, of relevance in connection with quantum quenches of
the XXZ quantum spin chain \cite{ADSV-16,S-17,CDV-18,S-20}.

The primary ingredients in the derivation of the multiple integral
representation \eqref{nefp_final} are: \textit{i)} the evaluation of
two alternative representations for the components of the off-shell
Bethe states $\psitop{r_1,\dots,r_s}$ and $\psibot{r_1,\dots,r_s}$,
that are essentially different from the longly known `coordinate
wavefunction' representation, and \textit{ii)} the symmetrization
relation \eqref{cantini}. Hopefully, such ingredients could turn
useful in the derivation of integral representations for more advanced
correlation functions, such as polarization.

The availaibility of two distinct integral representations for the
same quantity $F_N^{(r,s)}$ raises the natural question of their
mutual relation.  Consider for example the quantity $F_N^{(2,1)}$.
Evaluating it by means of \eqref{efpMIR2}, with $r=2$, $s=1$, or
\eqref{nefp_final}, with $s=1$, $n=1$, and equating the two results,
we get:
\begin{equation}
  h_N'(0)=[t^2 +(1-2\Delta t +t^2)h_{N-1}'(1)]h_N(0)
\end{equation}
relating the \textit{first} derivatives of function $h_N(z)$,
$h_{N-1}(z)$, evaluated at $z=0$, and $z=1$. Similarly, considering
$F_N^{(3,2)}$, one obtain an identity relating the \textit{second}
derivatives of function $h_N(z)$, $h_{N-1}(z)$, $h_{N-2}(z)$,
evaluated at $z=0$, and $z=1$. Remarkably, the coefficients appearing
in such relations do not depend on $N$. The game can be played
further, but the calculations become quite bulky very soon. The
existence of such hierachy of identities hints at some nontrivial
functional identity for $h_N(z)$, which is essentially Izergin-Korepin
partition function with just one inhomogeneity.


\section*{Acknowledgments}

We are indebted to N. Bogoliubov, L. Cantini, L. Petrov,
A. Sportiello, J.-M. St\'ephan, J. Viti, and P. Zinn-Justin, for
stimulating discussions at various stages of this work. The first
author (FC) acknowledges partial support from MIUR, PRIN grant
2017E44HRF on ``Low-dimensional quantum systems: theory, experiments
and simulations''. The third author (AGP) acknowledges partial support
from the Russian Science Foundation, grant 21-11-00141, and from INFN,
Sezione di Firenze.

\appendix

\section{`Coordinate wavefunction' representation}\label{app.wave}

Here we consider an alternative representation for
$\psitop{r_1,\dots,r_s}$ ensuing from the equivalence of the algebraic
and coordinate versions of Bethe ansatz. This equivalence was first
explicitly proved, as a side result, in \cite{IKR-87} (see appendix D
therein); see also book \cite{KBI-93}, Chap.~VII.  

In the case where all spectral parameters $\lambda_1,\dots,\lambda_N$
are taken to the same value $\lambda$, with the remaining spectral
parameters, $\nu_1,\dots,\nu_s$ left free, it reads
\begin{multline}\label{oldrep}
\psitop{r_1,\dots,r_s}=c^s \prod_{k=1}^s \left[a(\lambda,\nu_k)\right]^{N-1}
\prod_{1\leq j<k\leq s}\frac{1}{t_k-t_j}
\\
\times
\sum_{\sigma} (-1)^{[\sigma]}
\prod_{j=1}^s t_{\sigma_j}^{r_j-1}\prod_{1\leq j < k\leq s}
(t_{\sigma_j}t_{\sigma_k}-2\Delta t_{\sigma_j}+1),
\end{multline}
where
the sum is taken over the permutations $\sigma:1,\dots, s\mapsto
\sigma_1,\dots,\sigma_s$, with $[\sigma]$ denoting the parity of
$\sigma$, and 
\begin{equation}\label{t_k}
t_k=\frac{b(\lambda,\nu_k)}{a(\lambda,\nu_k)}.
\end{equation}
Clearly, formula \eqref{oldrep}, modulo the antisymmetric factor
$\prod_{1\leq j<k\leq s}(t_k-t_j)^{-1}$, coincides with the
$s$-particle coordinate Bethe ansatz wavefunction. Representation
\eqref{oldrep} has been discussed in details in various contexts,
especially in connection with the theory of symmetric polynomials,
see, e.g., \cite{M-17,B-17,BP-18,P-20}.

We need now to perform the homogeneous limit $\nu_1,\dots,\nu_s\to 0$,
which corresponds to sending $t_1,\dots,t_s\to t$. For this
purpose we implement the procedure explained in section 4.1.  Setting
$t_j=t w_j = t (1+\xi_j)$, we write
\begin{multline}\label{oldrep_det}
\psitop{r_1,\dots,r_s}=
c^s \prod_{k=1}^s \left[a(\lambda,\nu_k)\right]^{N-1}
\prod_{j=1}^s t^{r_j-j}
\prod_{1\leq j<k\leq s}\frac{1}{\xi_k-\xi_j}
\\ \times
\begin{vmatrix}
\exp(\xi_1\partial_{w_1}) & \dots & \exp(\xi_1\partial_{w_s})\\
\hdotsfor{3} \\
\exp(\xi_s\partial_{w_1}) & \dots & \exp(\xi_s\partial_{w_s})
\end{vmatrix}
\prod_{j=1}^s w_j^{r_j-1}
\\ \times
\prod_{1\leq j<k\leq s}(t^2 w_j w_k -2\Delta t w_j +1)\Bigg|_{w_1,\ldots,w_s=1}.
\end{multline}
This expression is still for the inhomogeneous model; it represents an
equivalent way of writing the multiple sum in \eqref{oldrep}.

Let us now perform the limit $\xi_1,\dots,\xi_s\to 0$. We readily
get
\begin{multline}\label{oldrep_hom}
  \psitop{r_1,\dots,r_s}=\frac{c^s a^{s(N-1)}}{\prod_{j=1}^{s-1} j!}
  \prod_{j=1}^s t^{r_j-j}
 \prod_{1\leq j<k\leq s}(\partial_{w_k}-\partial_{w_j})
\prod_{j=1}^s w_j^{r_j-1}
\\ \times
\prod_{1\leq j<k\leq s}
(t^2 w_j w_k -2\Delta t w_j +1)\Bigg|_{w_1,\ldots,w_s=1}.
\end{multline}
Representation \eqref{oldrep_hom} can now easily be turned into a
multiple integral representation by reexpressing the values of
derivatives as residues (or, equivalently, by using relation
\eqref{eqapp6} to replace sums with integrals in \eqref{oldrep} and
performing the homogeneous limit). We obtain
\begin{multline}\label{oldrep_mir_app}
\psitop{r_1,\dots,r_s}=c^s a^{s(N-1)}
\prod_{j=1}^s t^{r_j-j}
\oint_{C_1}^{} \cdots \oint_{C_1}^{} \prod_{j=1}^s \frac{w_j^{r_j-1}}{(w_j-1)^s}\\
\times
\prod_{1\leq j<k\leq s} \left[(w_j-w_k)(t^2 w_j w_k -2\Delta t w_j +1)\right]
\frac{\rmd^s w}{(2\pi \rmi)^s},
\end{multline}
thus reproducing \eqref{oldrep_mir}. It is worth to mention a rather
nontrivial property of representation \eqref{oldrep_mir_app}: if the
integrand is multiplied by a generic function $f(w_1,\dots,w_s)$, then the
result of the multiple integration is unaltered (modulo a trivial
prefactor $f(1,\dots,1)$) provided that $f(w_1,\dots,w_s)$ is
symmetric in all its variables, regular and nonvanishing in the
vicinity of $w_1=\dots = w_s=1$.  This property follows simply from
the fact that in representation \eqref{oldrep} an arbitrary symmetric
function of $t_1,\dots,t_s$ can be moved into or out of the sum.

Besides resorting to the equivalence of the algebraic and coordinate
Bethe ansatz, formulae \eqref{oldrep} and \eqref{oldrep_hom} can be
also derived by other methods. For instance, one can start from the
vertical monodromy matrix formulation \eqref{defZ1} for
$\psitop{r_1,\dots,r_s}$, and use the techniques of paper
\cite{KMT-99} to evaluate the matrix element. We also mention that
expression \eqref{oldrep_hom}, in a slightly different form, and for
special values of parameters, $\Delta=\frac{1}{2}$ and $t=1$, has also
been derived in the context of enumerative combinatorics \cite{F-06}.

Let us conclude with a short comment on the relation between
representation \eqref{topderivation3}, derived in the present paper
and the `coordinate wavefunction' representation \eqref{oldrep}.  It
appears that the latter is given by the residue at the poles due to
functions $a(\zeta_j,\nu_k)$ in the denominator of the integrand of
representation \eqref{topderivation3}.  Indeed, let us replace in
expression \eqref{topderivation3} all contours $C_\lambda$ with 
simple closed \textit{clockwise} contours
$C_{\nu-\eta}:=C_{\nu_1-\eta}\cup \dots\cup C_{\nu_s-\eta}$; evaluating the
residues, we obtain
\begin{multline}\label{differentpoles}
\frac{\prod_{k=1}^s
  \left[a(\lambda,\nu_k)\right]^{N-1} }{ \prod_{1\leq j <k \leq s}
  d(\nu_j,\nu_k)}\, Z_s(\nu_1-\eta,\dots,\nu_s-\eta;\nu_1,\dots,\nu_s)
\\ \times \sum_{\sigma} (-1)^{[\sigma]} \prod_{j=1}^{s}
t_{\sigma_j}^{r_j-1} \prod_{1\leq j <
  k\leq s} \frac{1}{e(\nu_{\sigma_k},\nu_{\sigma_j})},
\end{multline}
with  $t_k$  defined by \eqref{t_k}.
To proceed further we need the following identity
\begin{equation}\label{Z=prod}
Z_s(\nu_1-\eta,\dots,\nu_s-\eta;\nu_1,\dots,\nu_s)
=
c^s \prod_{\substack{j,k=1\\ j\ne k}}^s e(\nu_j,\nu_k),
\end{equation}
which can be proved by repeated use of the recursion relation for the
inhomogeneous partition function obtained in \cite{K-82}. Using
identity \eqref{Z=prod} and reexpressing everything in terms of
$t_k$'s, we readily recover the `coordinate wavefunction'
representation \eqref{oldrep}.

We have thus seen that, in representation \eqref{topderivation3}, the
integration contours $C_\lambda$ can be deformed into $C_{\nu-\eta}$
without modifying the result of the integration.  Taking into account
the periodicity of the integrand in \eqref{topderivation3}, and its
behaviour at infinity, it follows that the contribution arising from
the poles due to functions $e(\zeta_k,\zeta_j)$ in
\eqref{topderivation3} vanishes. This does not come as a surprise,
since, as already observed, such poles are only apparent, being
compensated by corresponding zeroes of the integrand. While this fact
is evident in the inhomogeneous model, in the homogeneous limit we
somehow loose track of it, and its verification requires some work,
see discussion in section~\ref{sec.deforming} for details.

\section{Dual representations for the `top' and `bottom' partition functions}
\label{app.dual}

We collect here various results for $\psitop{r_1,\dots,r_s}$, whose
derivation is based on relation \eqref{DB1}. First of all it is
convenient to resort to the alternative description of a given
configuration of a row, namely in terms of the positions of the $N-s$
down arrows $\rbar_1,\dots,\rbar_{N-s}$, rather than of the $s$ up
arrows. We can now rewrite \eqref{defZ1} as
\begin{multline}\label{defZ1bis}
\psitop{r_1,\dots,r_s}=\bra{\Downarrow\subtop}
B\subtop(\lambda_N) \cdots
B\subtop(\lambda_{\rbar_{N-s}+1}) \,D\subtop(\lambda_{\rbar_{N-s}})\,
B\subtop(\lambda_{\rbar_{N-s}-1})
\\ \times\cdots
B\subtop(\lambda_{\rbar_1+1})\,D\subtop(\lambda_{\rbar_1})
\,B\subtop(\lambda_{\rbar_1-1})
\cdots B\subtop(\lambda_1)
\ket{\Uparrow\subtop}.
\end{multline}
To evaluate expression \eqref{defZ1bis}, we proceed similarly to what
we have done in section \ref{sec.Zbot} for $\psibot{r_1,\dots,r_s}$,
namely we use \eqref{DB1} to move all operators $D\subtop(\lambda)$ to
the right, and make them act on $\ket{\Uparrow\subtop}$, exploiting
relation \eqref{Avac}. Iterating $N-s$ commutation relation
\eqref{derivationDB}, act on the right on $\ket{\Uparrow\subtop}$, and
multiplying form the left with the state vector
$\bra{\Downarrow\subtop}B\subtop(\lambda_N)\cdots
B(\lambda_{\rbar_{N-s}+1})$, we obtain
\begin{multline}\label{derivation22}
\psitop{r_1,\dots,r_s}=
\frac{
\prod_{\alpha=1}^N\prod_{k=1}^s
a(\lambda_{\alpha},\nu_k)b(\lambda_\alpha,\nu_k)}
{\prod_{1\leq\alpha<\beta\leq N}d(\lambda_\beta,\lambda_\alpha)
\prod_{1\leq j<k\leq s} d(\nu_j,\nu_k)}
\\
\times\sum_{\alpha_1=1}^{\rbar_1}
\sum_{\substack{\alpha_2=1\\ \alpha_2\ne\alpha_1}}^{\rbar_2}
\cdots
\sum_{\substack{\alpha_{N-s}=1\\
\alpha_{N-s}\ne\alpha_1,\dots,\alpha_{N-s-1}}}^{\rbar_{N-s}}
(-1)^{\sum_{j=1}^{N-s}(N-\alpha_j)-\sum_{1\leq j<k\leq N-s} \chi(\alpha_j,\alpha_k)}
\\
\times
\prod_{j=1}^{N-s} \tilde{v}_{\rbar_j}(\lambda_{\alpha_j})
\prod_{1\leq j<k\leq N-s}\frac{1}{e(\lambda_{\alpha_k},\lambda_{\alpha_j})}
\det\caM_{[\alpha_1,\dots,\alpha_{N-s};s+1,\dots,N]}.
\end{multline}
The functions $d(\lambda,\lambda')$ and $e(\lambda,\lambda')$ are
defined in \eqref{dfunc} and \eqref{efunc}, respectively, and we have
introduced, by analogy with \eqref{vrs}, the function
\begin{equation}\label{vts}
  \tilde{v}_{\rbar}(\lambda):=
  \frac{\prod_{\alpha=\rbar +1}^N d(\lambda,\lambda_\alpha)
    \prod_{\alpha=1}^{\rbar-1} e(\lambda,\lambda_\alpha)}
       {\prod_{k=1}^s a(\lambda,\nu_k)}.
\end{equation}
The function $\chi(\alpha, \beta)$ is defined in \eqref{chiab}.  We
also recall that $\caM_{[\alpha_1,\dots,\alpha_{N-s};s+1,\dots,N]}$
denotes the $s\times s$ matrix obtained from matrix $\caM$, see
\eqref{matT}, by removing rows $\alpha_1,\dots,\alpha_{N-s}$, and the
last $N-s$ columns. The equivalence of representations
\eqref{derivation22} and \eqref{derivation2} for
$\psitop{r_1,\dots,r_s}$ implies the existence of nontrivial
identities, which can be recognized as generalizing to $s$-fold sums
some identities already discussed for the case of a single sum in
\cite{G-83} (see also \cite{BPZ-02}).

We wish to underline that expression \eqref{derivation22}, can
alternatively be obtained directly from the corresponding expression
\eqref{derivation2} for $\psibot{r_1,\dots,r_s}$, via the crossing
symmetry transformation \eqref{duality1}.  Similarly, the orthogonal
polynomial representation, analogue to \eqref{orthZ2}, can be obtained
by direct computation, from expression \eqref{derivation22}, or
alternatively, just exploiting the homogeneous version of `duality'
relation \eqref{duality1}.  Note that the implementation of this
relation on expression \eqref{orthZ2} requires reversing the sign of
variables $\eps_j$, and hence of the corresponding derivatives
$\partial_{\eps_j}$.

In either way, we readily get the following orthogonal polynomial
representation:
\begin{multline}\label{orthZ1}
\psitop{r_1,\dots,r_s}=
\frac{Z_N}{a^{\frac{(N-s)(N-s-3)}{2}}
  b^{\frac{(N-s)(N+s+1)}{2}}c^{N-s}} \cdot \prod_{j=1}^{N-s}
\left(\frac{b}{a}\right)^{\rbar_j}
\\ \times
\begin{vmatrix}
K_{s}(\partial_{\eps_1}) &\hdots & K_{s}(\partial_{\eps_{N-s}})\\
\hdotsfor{3} \\
K_{N-1}(\partial_{\eps_1}) &\hdots & K_{N-1}(\partial_{\eps_{N-s}})
\end{vmatrix}
\prod_{j=1}^{N-s}\left\{ \frac{[\tilde\omega(\eps_j)]^{N-\rbar_j}}{
     [1-\tilde\omega(\eps_j)]^{s}}
\left[\frac{\omega(\eps_j)}{\tilde\omega(\eps_j)}\right]^{N-s-j} \right\}
\\ \times
\prod_{1\leq j<k \leq N-s}^{} \frac{1}
{1-\tilde\omega(\eps_k)\omega(\eps_j)}
\Bigg|_{\eps_1,\ldots,\eps_{N-s}=0}.
\end{multline}
To verify that representations \eqref{orthZ2} and \eqref{orthZ1}
indeed satisfy the duality relation \eqref{duality1}, a crucial
property of orthogonal polynomials $K_n(x)$ is needed, namely that
under a crossing transformation, $K_n(x;\lambda) \mapsto
K_n(x;\pi-\lambda) = (-1)^n K_n(-x;\lambda)$, see \cite{CP-05c}. Thus,
under the crossing  symmetry transformation, one has
\begin{equation}
K_n(\partial_{\eps_j}) \mapsto (-1)^n
K_n(\partial_{\eps_j}).
\end{equation}
On the other hand, the change of the sign in $\eps_j$, together with the
substitution $\lambda\mapsto\pi-\lambda$ gives rises to the following
substitution rule:
\begin{equation}
\omega(\eps_j) \mapsto \tilde{\omega}(\eps_j).
\end{equation}

We recall that, analogously to the corresponding representation for
$\psibot{r_1,\dots,r_s}$, representation \eqref{orthZ1} for
$\psitop{r_1,\dots,r_s}$ is valid for arbitrary values of the
parameters of the model, independently of the regime.

As a result, we have for $\psitop{r_1,\dots,r_s}$
the following  multiple integral representation:
\begin{multline}\label{MIRZ1}
\psitop{r_1,\dots,r_s}= Z_N\frac{\prod_{j=1}^{N-s}t^{\rbar_j-j}}
{b^{(N-s)(N-1)}
c^{N-s}}
\oint_{C_0}^{} \cdots \oint_{C_0}^{}
\prod_{j=1}^{N-s} \frac{1}{z^{\rbar_j}_j}
\\ \times
\prod_{1\leq j <k \leq N-s}
\frac{t^2(z_k-z_j)}{z_jz_k-2 \Delta t z_j +t^2}
\,\tilde h_{N,N-s}(z_1,\dots,z_{N-s})
\frac{\rmd^{N-s}z}{(2\pi \rmi)^{N-s}}.
\end{multline}
Here function $\tilde h_{N,s}(z_1,\dots,z_{s})$ is the function
resulting from $h_{N,s}(z_1,\dots,z_s)$ under exchange of the weights
$a$ and $b$. Function $\tilde h_{N,s}(z_1,\dots,z_{s})$ is defined
simply by replacing $h_N(z)$ with $\tilde h_N(z)$ in expression
\eqref{hNs}, where $\tilde h_N(z):=z^{N-1}h_N(z^{-1})$. As a
consequence, we have
\begin{equation}
\tilde h_{N,s}(z_1,\dots,z_s)= z_1^{N-1}\cdots z_s^{N-1}
h_{N,s}(z_1^{-1},\dots,z_s^{-1}).
\end{equation}
Multiple integral representation \eqref{MIRZ1} can be derived along
the lines of representation \eqref{MIRZ2} using a `crossing symmetry
transformed' version of relation \eqref{claim}, that is
\begin{equation}
K_{N-1}(\partial_\eps)\, f(\tilde\omega(\eps))\Big|_{\eps=0}=
\frac{1}{2\pi \rmi}\oint_{C_0}^{} \frac{(1- z)^{N-1}}{z^N} \tilde h_N(z)
f(z)\, \rmd z,
\end{equation}
or also directly from representation \eqref{MIRZ2}, by simple
application of the set of rules: $s\leftrightarrow (N-s)$,
$a\leftrightarrow b$, $t\leftrightarrow 1/t$, and
$h_{N}(z)\leftrightarrow \tilde h_{N}(z)$, implementing the  crossing
symmetry transformation.

Finally, for the sake of completeness let us give the representation
for $\psibot{r_1,\dots,r_s}$ that would come out by applying the
commutation relation \eqref{derivationBA}, rather than
\eqref{derivationAB}, to the expression \eqref{defZ2}. This is the
analogue of representation \eqref{MIRZ1alt} for
$\psitop{r_1,\dots,r_s}$, and immediately follows from it, thanks to
the crossing symmetry,
\begin{multline}\label{MIRZ2alt}
\psibot{r_1,\dots,r_s}
=Z_sb^{s(N-s)}\prod_{j=1}^{N-s} t^{j-\rbar_j}
\oint_{C_1} \cdots \oint_{C_1}
\prod_{j=1}^{N-s}\frac{(w_j-2\Delta t +t^2)^{\rbar_j-1}}{(w_j-1)^{\rbar_j}}
\\ \times
\prod_{1\leq j <k \leq N-s}\frac{w_k-w_j}{w_j w_k-2\Delta t w_j +t^2}
\\ \times
{\tilde h}_{N-s,N-s}(w_1,\dots,w_{N-s})
\frac{\rmd^{N-s}w}{(2\pi\rmi)^{N-s}}.
\end{multline}
Representations \eqref{MIRZ1} and \eqref{MIRZ2alt}, just as all other
representations, are valid for arbitrary values of the parameters of
the model, independently of the regime.

\section{A remarkable identity}
\label{app.id}

We comment here on some remarkable identity stemming from the
identification of the two alternative representations resulting for
$\psitop{r_1,\dots,r_s}$ by applying the QISM machinery in the
`horizontal' or in the `vertical' direction. We focus on the situation
where all inhomogeneities are turned on.

On the one hand, we have the `coordinate wavefunction' representation,
whose homogeneous version has already been discussed in
appendix~\ref{app.wave}, and whose fully inhomogeneous version, where
all the spectral parameters $\lambda_1,\dots,\lambda_N$ and
$\nu_1,\dots,\nu_s$ are left free, takes the form \cite{B-71}:
\begin{multline}\label{oldrep_inhom}
\psitop{r_1,\dots,r_s}=c^s\prod_{\beta=1}^N\prod_{k=1}^s a(\lambda_\beta,\nu_k)
\prod_{1\leq j<k\leq s}\frac{1}{d(\nu_j,\nu_k)}
\\
\times
\sum_{\sigma} (-1)^{[\sigma]}
\prod_{\beta_1=1}^{r_1-1}
\frac{b(\lambda_{\beta_1},\nu_{\sigma_1})}{a(\lambda_{\beta_1},\nu_{\sigma_1})}\dots
\prod_{\beta_s=1}^{r_s-1}
\frac{b(\lambda_{\beta_s},\nu_{\sigma_s})}{a(\lambda_{\beta_s},\nu_{\sigma_s})}
\\
\times\prod_{j=1}^s\frac{1}{a(\lambda_{r_j},\nu_{\sigma_j})}
\prod_{1\leq j<k\leq s}e(\nu_{\sigma_j},\nu_{\sigma_k}),
\end{multline}
see also \cite{B-73}, Eqs. (79) and (80), and discussion in
\cite{B-87}.

On the other hand, we have the alternative representation
\eqref{topderivation1bis}.  Identification of these two expressions
clearly implies the existence of some relation which, after
cancellation of some nonrelevant prefactors, reads:
\begin{multline}\label{big-id}
c^s\prod_{1\leq j<k\leq s}\frac{1}{d(\nu_j,\nu_k)}
\sum_{\sigma} (-1)^{[\sigma]}
\prod_{\beta_1=1}^{r_1-1}
\frac{b(\lambda_{\beta_1},\nu_{\sigma_1})}{a(\lambda_{\beta_1},\nu_{\sigma_1})}\dots
\prod_{\beta_s=1}^{r_s-1}
\frac{b(\lambda_{\beta_s},\nu_{\sigma_s})}{a(\lambda_{\beta_s},\nu_{\sigma_s})}
\\
\shoveright{
\times  \prod_{j=1}^s\frac{1}{a(\lambda_{r_j},\nu_{\sigma_j})}
\prod_{1\leq j<k\leq s}e(\nu_{\sigma_j},\nu_{\sigma_k})
=
}
\\
\shoveleft{
  =
\sum_{\alpha_1=1}^{r_1}
\sum_{\substack{\alpha_2=1\\ \alpha_2\ne\alpha_1}}^{r_2}
\cdots
\sum_{\substack{\alpha_s=1\\ \alpha_s\ne\alpha_1,\,\dots,\alpha_{s-1}}}^{r_s}
\prod_{j,k=1}^s\frac{1}{a(\lambda_{\alpha_j},\nu_k)}\ 
Z_{s}\left(\lambda_{\alpha_1},\dots,\lambda_{\alpha_s};\nu_1,\dots,\nu_s \right)}
\\
\times
\prod_{\substack{\beta_1=1\\ \beta_1 \ne\alpha_1}}^{r_1}
\frac{1}{d(\lambda_{\alpha_1},\lambda_{\beta_1})}
\prod_{\substack{\beta_2=1\\ \beta_2 \ne\alpha_1,\alpha_2}}^{r_2}
\frac{1}{d(\lambda_{\alpha_2},\lambda_{\beta_2})} \cdots
\prod_{\substack{\beta_s=1\\ \beta_s \ne\alpha_1,\dots,\alpha_s}}^{r_s}
\frac{1}{d(\lambda_{\alpha_s},\lambda_{\beta_s})}
\\
\times
\prod_{\beta_1=1}^{r_1-1} e(\lambda_{\alpha_1},\lambda_{\beta_1})\cdots
\prod_{\beta_s=1}^{r_s-1} e(\lambda_{\alpha_s},\lambda_{\beta_s})
\prod_{1\leq j<k \leq s} \frac{1}{e(\lambda_{\alpha_k},\lambda_{\alpha_j})}.
\end{multline}
The identity holds true by construction. However, a simple and direct
proof can be also given, by reexpressing both sides of \eqref{big-id}
as contour integrals, and then proceeding along the lines of the
discussion provided at the end of appendix~\ref{app.wave}.

Let us consider some particular instance of the identity \eqref{big-id}.
To start with, let us  set $s=1$. The above relation 
reduces to
\begin{equation}
  \frac{\prod_{\beta=1}^{r-1}b(\lambda_\beta,\nu)}
       {\prod_{\beta=1}^{r}a(\lambda_\beta,\nu)}
  =\sum_{\alpha=1}^{r}\frac{1}{a(\lambda_\alpha,\nu)}
  \frac{\prod_{\beta=1}^{{r-1}}e(\lambda_\alpha,\lambda_\beta)}
       {\prod_{\substack{\beta=1\\\beta\ne \alpha}}^r d(\lambda_\alpha,\lambda_\beta)},
\end{equation}
which is a particular case of equation (45) of \cite{BPZ-02}, see also
\cite{G-83}, and readily implies the identity between
\eqref{topderivation1bis} at $s=1$ and \eqref{ztops=1}.

Let us now turn to investigate the identity \eqref{big-id} for generic
values of $s$, but with the lattice coordinates chosen to be $r_j=j$,
$j=1,\dots,s$. We obtain:
\begin{multline}\label{C4}
c^s\prod_{1\leq j<k\leq s}\frac{1}{d(\nu_j,\nu_k)}
\sum_{\sigma}(-1)^{[\sigma]} \prod_{1\leq j<k\leq s}
a(\lambda_k,\nu_{\sigma_j}) b(\lambda_j,\nu_{\sigma_k})
e(\nu_{\sigma_j},\nu_{\sigma_k})
\\
=Z_s(\lambda_1,\dots,\lambda_s;\nu_1,\dots,\nu_s),
\end{multline}
which reproduces essentially the antisymmetrization relation \eqref{KMST},
that is Proposition C.1 in \cite{KMST-02}.

Let us now consider the identity \eqref{big-id} with the first $s-1$
lattice coordinates specialized to the values $r_j=j$,
$j=1,\dots,s-1$, while the last one is left generic, $r_s=:r$. 
This configuration of $r_j$'s appears 
in the context of the so-called `tangent method' \cite{CS-16}. 
After some calculation, and use of relation \eqref{C4}, we obtain:
\begin{multline}\label{tangent}
  \sum_{l=1}^s\prod_{\substack{j=1\\ j\ne l}}^s\frac{e(\nu_j,\nu_l)}{d(\nu_j,\nu_l)}
    \prod_{\substack{k=1\\ k\ne l}}^s a(\lambda_s,\nu_k)
      \prod_{\beta=1}^{r-1}b(\lambda_\beta,\nu_l)
      \prod_{\beta=s+1}^r\frac{1}{a(\lambda_\beta,\nu_l)}
      Z_{s-1}[\lambda_s;\nu_l]
\\
=
\sum_{\alpha=s}^r\prod_{k=1}^s\frac{a(\lambda_s,\nu_k)}{a(\lambda_\alpha,\nu_k)}
    \prod_{\substack{\beta=s \\ \beta\ne\alpha}}^r
       \frac{e(\lambda_\alpha,\lambda_\beta)}{d(\lambda_\alpha,\lambda_\beta)}
       \frac{1}{e(\lambda_\alpha,\lambda_r)}
       Z_s(\lambda_1,\dots,\lambda_{s-1},\lambda_\alpha;\nu_1,\dots,\nu_s).
\end{multline}
Note that, unlike both sides of relation \eqref{C4}, the left- and
right-hand sides of \eqref{tangent} clearly reflect in their structures
the application of the fundamental commutation relations
\eqref{AA}-\eqref{BD1} in the `horizontal' and `vertical'
directions, respectively.

\bibliography{intreps_bib}
\end{document}